\documentclass[twocolumn]{aastex631}
\usepackage{apjfonts}
\usepackage{graphics,graphicx,subfigure,float,color,amsmath,natbib}
\usepackage{multirow}
\usepackage{hyperref}
\usepackage{listings}
\usepackage{color}
\usepackage[table]{xcolor}

\definecolor{Gray}{gray}{0.9}

\newcommand{\W}{$\lambda$}

\newcommand{\CH}[1]{\colhead{#1}}

\newcommand{\mcV}[1]{\multicolumn{5}{c}{#1}}

\begin{document}

\shortauthors{Trevino et al.}

\title{Recovering Ionizing Photon Escape and Galaxy Scaling Relations in the LzLCS via \ion{Si}{2} and \ion{C}{2} Absorption Lines and Mock Spectra from a Radiation–Hydrodynamic Simulation}

\author[0009-0008-9637-757X]{John Trevino}
\affiliation{Department of Astronomy, The University of Texas at Austin, 2515 Speedway, Stop C1400, Austin, TX 78712, USA}

\author[0000-0002-5659-4974]{Simon Gazagnes}
\affiliation{Scuperta, Intelligent Embedded Systems Department, 9741AS, Groningen,  The Netherlands}

\author[0000-0002-4153-053X]{Danielle A. Berg}
\affiliation{Department of Astronomy, The University of Texas at Austin, 2515 Speedway, Stop C1400, Austin, TX 78712, USA}
\affiliation{Cosmic Frontier Center, The University of Texas at Austin, Austin, TX 78712, USA}

\author[0000-0002-8809-4608]{Kaelee S. Parker}
\affiliation{Department of Astronomy, The University of Texas at Austin, 2515 Speedway, Stop C1400, Austin, TX 78712, USA}
\affiliation{Cosmic Frontier Center, The University of Texas at Austin, Austin, TX 78712, USA}

\author[0000-0003-0595-9483]{Valentin Mauerhofer}
\affiliation{Kapteyn Astronomical Institute, University of Groningen, P.O Box 800, 9700 AV Groningen, The Netherlands}

\author[0000-0003-1609-7911]{Jeremy Blaizot}
\affiliation{Centre de Recherche Astrophysique de Lyon UMR5574, F-69230, Saint-Genis-Laval, France}

\author[0000-0002-2201-1865]{Anne Verhamme}
\affiliation{Department of Astronomy, University of Geneva, 51 Chemin Pegasi, 1290 Versoix, Switzerland}

\author[0000-0002-0159-2613]{Sophia R. Flury}
\affiliation{Institute for Astronomy, University of Edinburgh, Royal Observatory, Edinburgh, EH9 3HJ, UK}

\author[0000-0002-6790-5125]{Anne E. Jaskot}
\affiliation{Department of Physics and Astronomy, Williams College, Williamstown, MA 01267, USA}

\author[0000-0001-8419-3062]{Alberto Saldana-Lopez}
\affiliation{Stockholm University, Department of Astronomy and Oskar Klein Centre for Physics, AlbaNova University Centre, SE-10691, Stockholm, Sweden}

\begin{abstract}
In this work, we use a radiation-hydrodynamic simulation of a single $\sim 10^9\ M_\odot$ virtual galaxy to study \ion{Si}{2} and \ion{C}{2} line profiles seen in stacked HST/COS spectra of 58 galaxies from the LzLCS+ sample. We compare stacks across three mass bins ($M_\star \le 10^8$, $10^8$--$10^9$, and $\ge 10^9\ M_\odot$) and three stacking methods (mean, median, weighted averaged) to a library of 22,500 mock spectra. We investigate if the simulation can accurately mimic real gas features,  reveal clear trends with galaxy properties, and provide indirect estimates of the ionizing escape fraction ($f_{\rm esc}$). We find reasonable agreement between simulated and observed profiles ($\chi^2 < 1$) across all mass regimes. Notably, extracting line properties such as EW and $\text{R}_f$ from best-fit mock profiles provides a robust alternative to direct empirical trends, particularly in the low-S/N regime where noise frequently biases results. The simulation-based LIS features, although derived from a single virtual object, exhibit clear correlations with $M_\star$, SFR, and $f_{\rm esc}$, mirroring established empirical scaling relations. We find that the best-matching mock spectra predominantly originate from simulation time steps corresponding to peak UV luminosity and intense starburst phases, suggesting these active periods generate the ISM diversity observed in star-forming galaxies. Finally, simulation-based estimates ($f_{\rm esc}^{\rm virtual}$) reproduce the observed mass-dependent trends in $f_{\rm esc}$ and are in close agreement with the average $f_{\rm esc}$ of the generated stacks. This simulation-based framework establishes a relevant methodology for interpreting spectroscopic observations, including inferring $f_{\rm esc}$ and characterizing physical scaling relations, in high-redshift galaxies from the Epoch of Reionization.

\end{abstract}



\section{Introduction}\label{sec1}

The Epoch of Reionization (EoR) represents a critical stage in cosmic history, when the intergalactic medium (IGM) transitioned from neutral to ionized. 
Identifying the sources and mechanisms that drove this process is central to our understanding of galaxy formation and the growth of large-scale structure. 
While faint, star-forming galaxies are widely regarded as the dominant contributors to reionization \citep[e.g.,][]{robertson15, finkelstein19}, some studies suggest that active galactic nuclei (AGN) may have provided a non-negligible contribution, particularly at the bright end of the luminosity function \citep{dayal24, grazian24}. 
One of the outstanding challenges in extragalactic astrophysics is determining how their ionizing radiation escaped into the IGM.

At high redshift ($z\gtrsim4$), direct detections of hydrogen-ionizing photons (Lyman continuum, LyC) are hindered by the increasing opacity of the IGM \citep[e.g.,][]{inoue14}. 
This limitation necessitates the use of indirect indicators to constrain LyC escape fractions. 
Indirect methods, however, require calibration against lower-redshift analogues where LyC leakage can be measured directly. 
The Low-Redshift Lyman Continuum Survey \citep[LzLCS;][]{flury22a,flury22b} provides the largest uniform sample of confirmed LyC emitters in the nearby ($z\sim0$) universe, offering an ideal testbed for such calibrations.

A particularly promising approach to inferring LyC escape fractions relies on far-ultraviolet (FUV) absorption features from LIS metals, including C$^+$ and Si$^+$. 
With ionization potentials comparable to that of neutral hydrogen (C$^+$: 11.26–24.38 eV; Si$^+$: 8.15–16.34 eV), these species trace the neutral gas that absorbs LyC escape. 
Prominent LIS transitions, such as \ion{Si}{2} \W\W1260,1265 and \ion{C}{2} \W1334, probe the covering fraction, kinematics, and porosity of the ISM—properties intimately tied to feedback from massive stars and supernovae \citep[e.g.,][]{heckman11,carr23,xu23,parker24,flury25}. 
Yet, translating these absorption-line profiles into quantitative predictions of LyC escape remains challenging, as they are sensitive to geometry, radiative transfer effects, and viewing angle.

Radiation-hydrodynamic (RHD) simulations provide a powerful framework to bridge the gap between LIS absorption line diagnostics and observations. 
By modeling stellar feedback, ionizing photon transport, and synthetic spectra, RHD simulations link the physical conditions of the ISM to observable absorption features \citep[e.g.,][]{kimm14,mauerhofer21, mauherhofer26}. 
For example, \citet{gazagnes23, gazagnes24} successfully used mock spectrum from a single RHD simulation to interpret LIS absorption two distinct sample: (i) high continuum signal-to-noise (S/N$\sim6.4$), high resolution ($R\sim15,000$) spectra from the COS Legacy Spectroscopic SurveY \citep[CLASSY;][]{berg22, james22}, and (ii) galaxies at 3.35 $<$ $z$ $<$ 3.95, drawn from the ultra-deep VANDELS spectroscopic survey \citep{McLure2018_vandels, Pentericci2018_vandels, Garilli2021_vandels}, demonstrating the potential of such simulation in reproducing the observed line profiles of star-forming galaxies.

The behavior of low-ionization absorption lines in the LzLCS sample was previously explored in several studies. In particular, \citet{flury25} performed a detailed stacking analysis of the LzLCS$+$ galaxies to investigate how LIS absorption properties relate to ionizing escape fraction and galaxy properties. Their study demonstrated that galaxies with higher LyC escape fractions tend to show weaker LIS absorption and higher residual flux, consistent with reduced neutral gas covering fractions. \citet{flury25} also examined the methodological effects of stacking choices, spectral resolution, and the COS line spread function on measured absorption properties. 

The present work builds upon their work but focuses on comparison with a RHD simulation. Specifically, we compare the LIS absorption lines from the RHD simulation from \citet{mauerhofer21} to the LzLCS observations which is a unique sample because it contains direct LyC escape measurements. The LzLCS spectra have lower resolution ($R\sim1050$) than CLASSY ($R\sim3000$) but higher than VANDELS ($R\sim600$), and are representative of galaxy samples across a broad range of redshifts \citep{izotov21}. In addition to the moderate spectral resolution of the LzLCS+ spectra, the individual spectra are intrinsically noisy, with the non-ionizing UV continuum typically detected at only $\sim1\sigma$ per resolution element near rest-frame 1100 \AA, a region where the stellar continuum is otherwise relatively smooth and free of strong spectral features \citep{flury25}. \citet{gazagnes23} showed that moderate spectral resolution alone does not preclude a meaningful comparison between observations and simulated spectra. For the LzLCS+, however, the characteristic S/N of the individual LIS absorption lines is low ($<1$), which prevents a reliable one-to-one comparison between individual observed and simulated spectra. This work therefore stacks the spectra in stellar mass bins and compare the resulting composites with mock observations generated using postprocessing of the virtual galaxy. Our analysis (1) investigates whether simulated absorption profiles reproduce the observed LzLCS stacked spectra, (2) compares the ISM conditions and galaxy properties that shape low-ionization absorption features in the observations and in the simulation, and (3) evaluates the predictive power of using a RHD simulation combined to low-resolution \ion{Si}{2} and \ion{C}{2} profiles as an alternative diagnostic of the LyC escape. 

The paper is organized as follows. 
Section~\ref{sec2} presents the LzLCS$+$ dataset (\S~\ref{sec2.1}) and the RHD simulation (\S~\ref{sec3}). Section~\ref{sec4} matches the observed stacked spectra with the simulated mock LIS profiles and explores the physical origin of their resemblance. Section~\ref{sec5} places our results in the context of previous studies targeting LzLCS LIS line profiles, as well as works comparing RHD simulations to observations. We also discuss the relevance of these findings for high-redshift studies, along with the limitations of this work. Finally, Section~\ref{sec6} summarizes our main conclusions.
Throughout this work, we adopt a flat $\Lambda$CDM cosmology with $H_{0} = 70$km/s/Mpc, $\Omega_{m} = 0.3$, and $\Omega_{\Lambda} = 0.7$. 
All velocities are reported in the rest frame of the systemic redshift and in units of km s$^{-1}$. 
Metallicities are expressed relative to the solar oxygen abundance following \citet{asplund21}, where $12 + \log(\mathrm{O/H})_{\odot} = 8.69$.
When interpreting the simulation results, we assume a scaled-solar abundance pattern and 
stellar population synthesis consistent with the BPASS~v2.0 models adopted in the RHD simulation. 
These assumptions provide a consistent framework for comparing observed and simulated metal-line absorption features across the LzLCS$+$ sample.

\section{Data}\label{sec2}

\subsection{The LzLCS Sample}\label{sec2.1}

\begin{figure*}[ht]
    \centering
	\includegraphics[width=\linewidth]{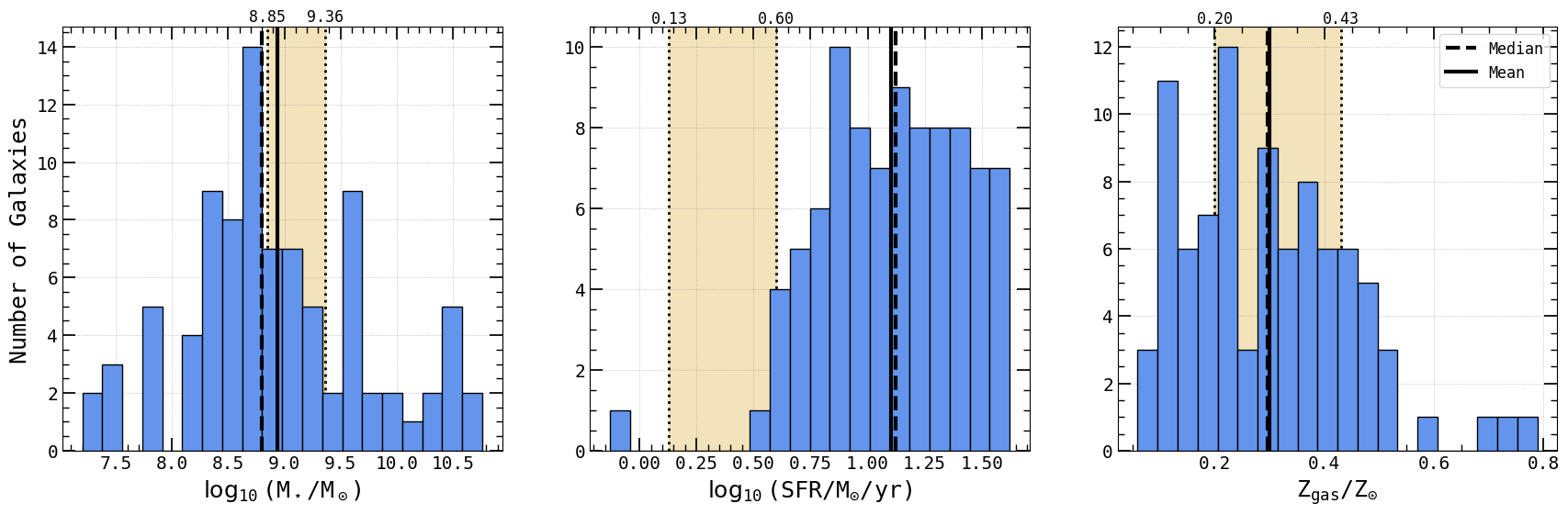}
\caption{Distributions of stellar mass (left), star formation rate (center), and gas-phase metallicity (right) for galaxies in the LzLCS$+$ sample. 
The dashed and solid black lines show the sample median and mean values, respectively. The shaded golden regions denote the parameter space spanned by the RHD simulation,
where the dotted vertical lines mark the bounds of these ranges. The SFR values correspond to averages over the past 100 Myr, following \cite{mauerhofer21}.
The observed LzLCS$+$ sample spans over three orders of magnitude in stellar mass ($10^{7.2}-10^{10.8}\ M_\odot$) and over two orders of magnitude in SFR ($0.74-42\ M_\odot$ yr$^{-1}$), the latter enhanced relative to the $z\sim0$ star-forming main sequence, with metallicities ranging between $0.06-0.79\ Z_\odot$.}
\label{fig:hist of M, SFR, Z}
\end{figure*}


The \textit{Low-Redshift Lyman Continuum Survey} (LzLCS) is a large Hubble Space Telescope Cosmic Origins Spectrograph (HST/COS) program (GO 15626, PI: Jaskot) that provides the most extensive spectroscopic investigation of Lyman continuum (LyC) emission at low redshift to date \citep{flury22a,flury22b}. 
The survey targeted 66 star-forming galaxies at $z\sim0.2-0.4$ with the COS G140L grating, which includes rest-frame coverage that extends blue-wards of the hydrogen Lyman limit ($<912$ \AA). 
This direct access to the LyC wavelengths allows for robust measurements of ionizing photon escape fractions.
The escape fraction is then derived by comparing the observed LyC flux to the intrinsic LyC production predicted from stellar population synthesis (SSP) modeling of the galaxies' spectral energy distributions. In particular, \citet{flury22a} and \citet{saldana-lopez22} model the intrinsic ionizing flux using SED fits constrained by the non-ionizing UV continuum and nebular emission, allowing for an estimate of the absolute LyC escape fraction ($f_{\rm esc}^{\rm abs}$). This approach accounts for the underlying stellar populations and star-formation histories, enabling a physically motivated conversion between the observed LyC flux and the total ionizing photon budget.
LyC detection in LzLCS was performed by carefully modeling and subtracting geocoronal contamination, estimating background levels with multiple extraction windows, and quantifying flux recovery below the Lyman limit \citep{flury22a}.  

The LzLCS sample was selected to span a broad range of galaxy properties that are critical for understanding LyC escape. Specifically, targets were drawn from SDSS and GALEX catalogs to include compact, star-forming galaxies at $z \sim 0.2$--$0.4$ with strong nebular emission lines, often exhibiting elevated \ion{O}{3}/\ion{O}{2} ratios ($\mathrm{O}_{32}$), high star-formation rate surface densities, and relatively low metallicities---properties that have been empirically associated with enhanced LyC escape \citep{flury22a}.
The galaxies cover stellar masses of $M_\star\sim10^{7.2}-10^{10.8}M_\odot$, star formation rates of $\sim0.74-42\ M_\odot$ yr$^{-1}$, and gas-phase metallicities of $12+\log(\text{O/H})\sim7.46-8.59$. 
Figure \ref{fig:hist of M, SFR, Z} illustrates these property distributions, highlighting the wide dynamic range in $M_\star$, star formation rate (SFR), and $Z_\text{gas}/Z_\odot$ across the sample with dashed and solid vertical lines marking the median and mean values, respectively. 
The LzLCS sample selection favored compact UV-bright galaxies, which are expected to be analogs of high-redshift reionization-era systems \citep{mascia24}. 
By combining these selection criteria with uniform COS spectroscopy, the LzLCS enables a controlled study of how galaxy-scale parameters regulate LyC leakage.  

In addition to the 66 core targets, the LzLCS+ sample incorporates 23 previously observed LyC emitter candidates at $z\lesssim0.5$ from \cite{izotov16a, izotov16b, izotov18a, izotov18b, izotov21, wang19}, extending the dynamic range of galaxy properties and increasing the number of confirmed LyC leakers. 
Together, LzLCS+ represent the largest and most homogeneous dataset of spectroscopically confirmed LyC detections in the low-redshift universe.

Previous analyses of the LzLCS$+$ galaxies generally favor bursty star-formation histories rather than smooth, continuous star formation. The sample is also dominated by compact UV-bright systems, with typical UV half-light radii of approximately $0.3 - 0.6$ kpc, although a small number of objects extend to radii of approximately $2$ kpc. These compact morphologies and burst-dominated star-formation histories are characteristic of the high-specific-SFR systems selected as low-redshift analogs of galaxies in the reionization era.

The LzLCS$+$ sample is particularly well-suited for the purposes of this work. 
Its combination of direct LyC coverage, well-characterized detections, and diversity in galaxy physical conditions provides an unparalleled opportunity to test correlations between ionizing photon escape and other rest-frame UV diagnostics. 
Moreover, as a statistically significant and uniform dataset, the LzLCS$+$ offers an essential low-redshift analog to reionization-era galaxies, making it an ideal benchmark for interpreting rest-UV absorption diagnostics. Future deep spectroscopic observations of gravitationally lensed galaxies at higher redshift will provide a powerful complementary test of these diagnostics. For instance, recent JWST programs targeting strongly magnified systems with NIRSpec IFU spectroscopy are designed to obtain spatially resolved rest-frame UV and optical spectra of faint galaxies, enabling detailed measurements of outflow kinematics, column densities, and gas structure that remain difficult to access in unlensed high-redshift galaxies \citep{xu24}.  

\begin{deluxetable*}{lr|ccc|lccc}
\tablecaption{Properties of the LzLCS+ stacks used in this work.}
\tablehead{
\CH{} & \CH{} \vline & \CH{$\log$(M$_\star$)} & \CH{$\log$(SFR)}  & \CH{$f_{\mathrm{esc}}$} \vline & & \multicolumn{3}{c}{S/N} \\[-1ex] 
\cline{7-9}
\CH{Bins}   & \CH{\#} \vline & \CH{\footnotesize{(M$_\odot$)}}    & \CH{\footnotesize{(M$_\odot$ yr$^{-1}$)}} & & \CH{Line} \vline  & \CH{Mean Stack} & \CH{Median Stack} & \CH{W. Ave. Stack}\vline }

\startdata
\multirow{2}{*}{$M_\star<10^8$} & \multirow{2}{*}{9}  & \multirow{2}{*}{$7.57^{+0.23}_{-0.20}$} & \multirow{2}{*}{$0.92^{+0.09}_{-0.17}$} & \multirow{2}{*}{$0.19^{+0.17}_{-0.17}$} & \ion{Si}{2} \W1260 & 3.13 & 2.68 & 3.61 \\
& & & & & \ion{C}{2} \W1334 & 2.98 & 1.48 & 1.69\\
\cline{6-9}
\multirow{2}{*}{$10^8 < M_\star <10^{9}$} & \multirow{2}{*}{29} & \multirow{2}{*}{$8.61^{+0.29}_{-0.29}$} & \multirow{2}{*}{$1.08^{+0.29}_{-0.27}$} & \multirow{2}{*}{$0.11^{+0.02}_{-0.10}$} & \ion{Si}{2} \W1260 & 6.76 & 5.88 & 8.48 \\
& & & & & \ion{C}{2} \W1334 & 5.22 & 4.44 & 6.08 \\
\cline{6-9}
\multirow{2}{*}{$M_\star >10^{9}$}     & \multirow{2}{*}{20} & \multirow{2}{*}{$9.54^{+0.37}_{-0.40}$} & \multirow{2}{*}{$1.10^{+0.34}_{-0.24}$} & \multirow{2}{*}{$0.04^{+0.01}_{-0.03}$} & \ion{Si}{2} \W1260 & 6.89 & 5.82 & 7.19 \\
& & & & & \ion{C}{2} \W1334 & 4.49 & 4.70 & 5.23
\enddata
\tablecomments{Columns 1 and 2 list the three stellar mass bins of each stack and the the corresponding number of galaxies in each bin. 
Columns 3--5 list the median stellar mass, SFR, and escape fraction of the galaxies in each bin. The values for individual LzLCS+ galaxies come from \cite{flury22a} and \cite{flury22b}.
Columns 7--9 list the S/N measured in the continuum near the corresponding absorption line
for each of the three stacking methods used (see Section~\ref{sec2.2} for further details).
\label{table:stacking numbers}}
\end{deluxetable*}


\begin{figure*}[ht]
    \centering
	\includegraphics[width=\linewidth]{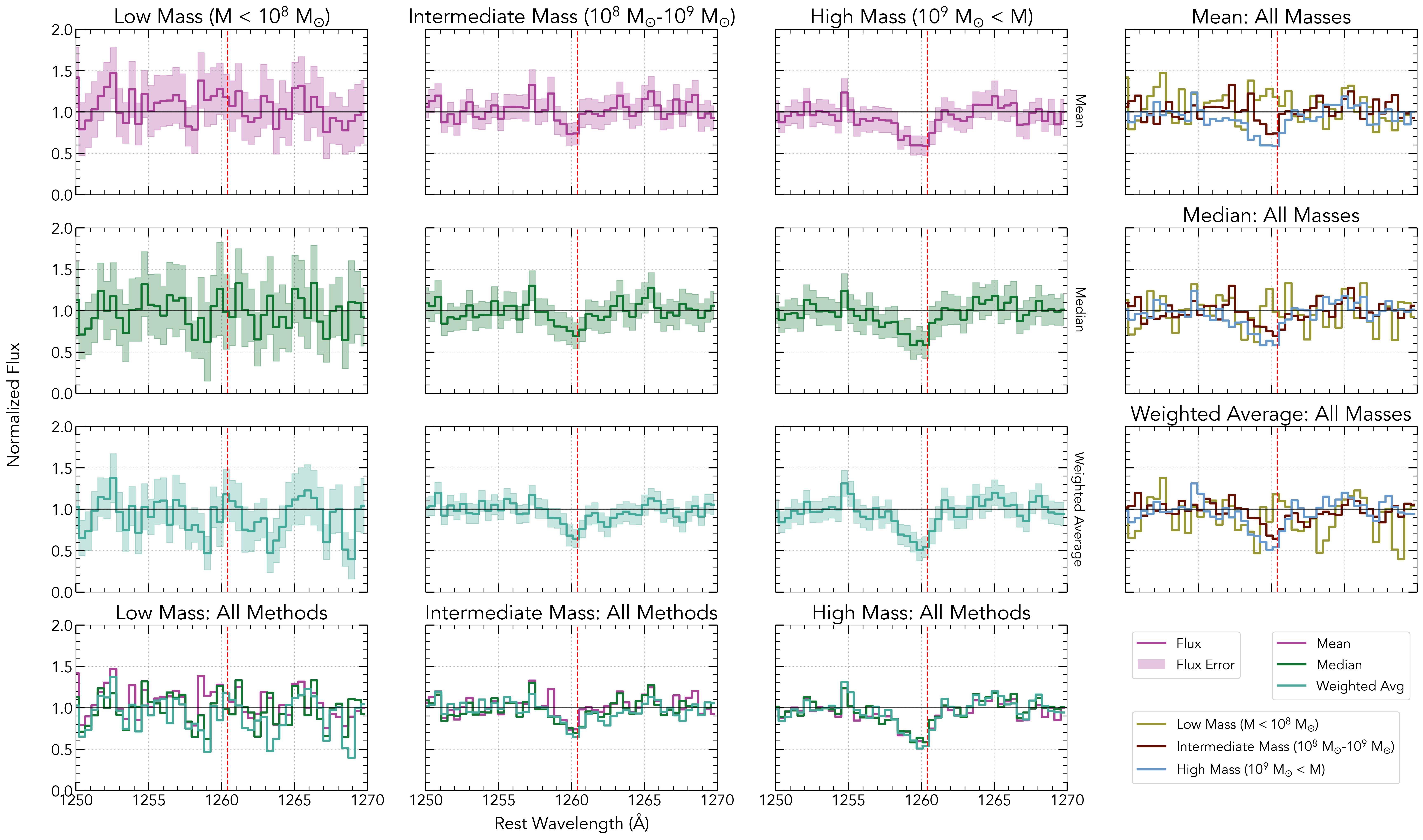}
\caption{LzLCS+ normalized spectral stacks of the \ion{Si}{2} $\lambda1260.42$ absorption feature. Each row shows one of the three stacking methods applied (mean, median, weighted average), while the columns correspond to stellar mass bins: low mass ($M_\star < 10^8 M_\odot$), intermediate mass ($10^8 < M_\star/M_\odot < 10^9$), and high mass ($M_\star > 10^9 M_\odot$). The fourth row overplots all three stacking methods for direct comparison within each mass bin, while the fourth column overplots all three mass bins for each stacking method. The vertical red dashed line marks the rest wavelength of the \ion{Si}{2} transition. All three stacking approaches provide very similar stacked profiles, with increasing \ion{Si}{2} absorption depth and width with mass.}
\label{fig:SiII Stack}
\end{figure*}

\begin{figure*}[ht]
    \centering
	\includegraphics[width=\linewidth]{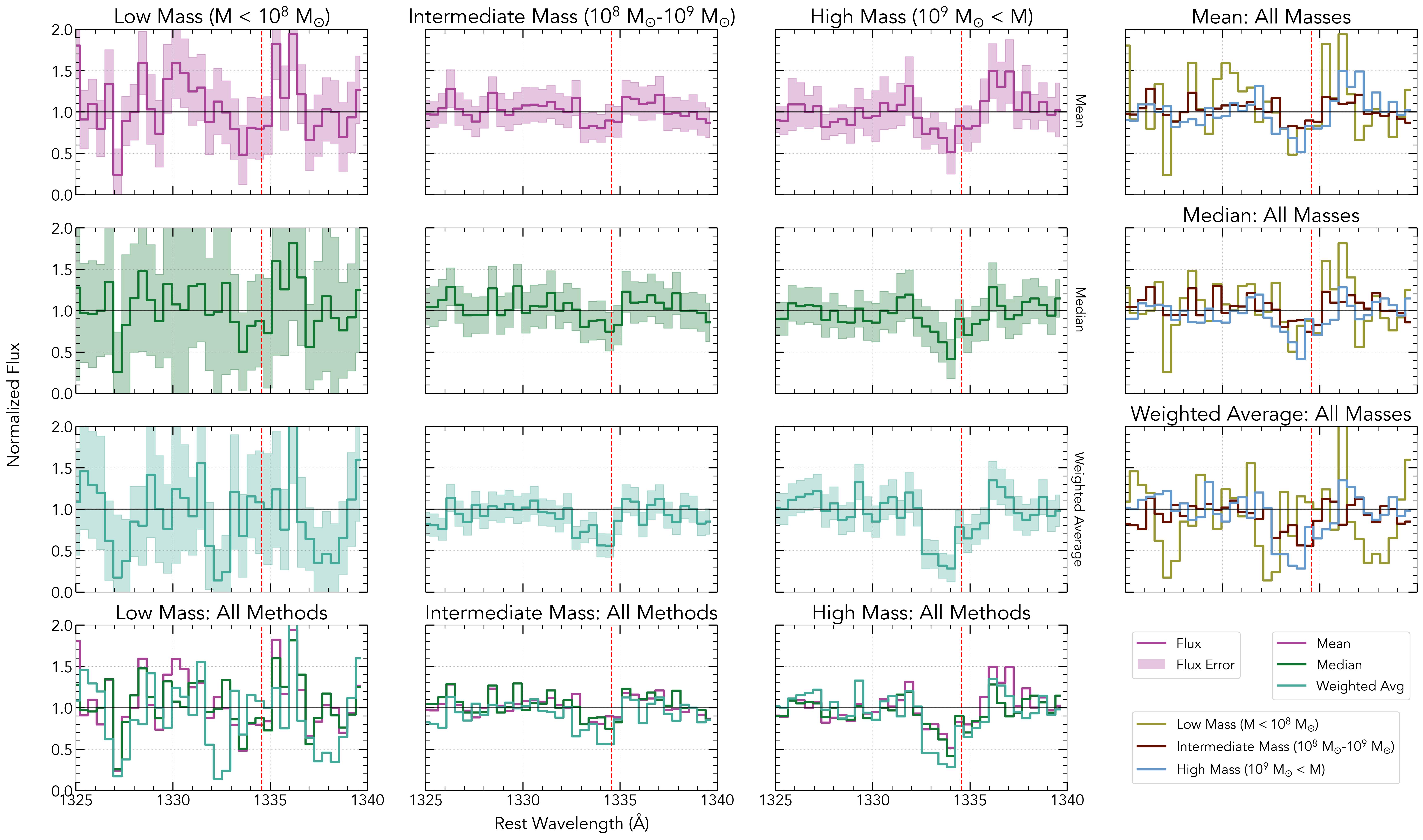}
\caption{Same as Figure~\ref{fig:SiII Stack}, but for the \ion{C}{2} $\lambda1334.57$ absorption feature. Each row corresponds to a stacking method, and the columns represent the same three stellar mass bins. The fourth row overplots the mean, median, and weighted-average stacks for direct visual comparison, while the fourth column overplots all three mass bins for each stacking method. The vertical red dashed line indicates the rest wavelength of the \ion{C}{2} transition. The strongest \ion{C}{2} features are seen in the intermediate- and high-mass bins, suggesting larger gas column density in more massive galaxies.}
\label{fig:CII Stack}
\end{figure*}

\subsubsection{Stacking Method}\label{sec2.2}

As mentioned in the introduction, the individual LzLCS+ spectra typically have S/N below $1\sigma$ around the \ion{Si}{2} \W1260 and \ion{C}{2} \W1334 ISM absorption features. Therefore, we chose to stack these spectral features using a methodology similar to that of \cite{flury25}. The increased signal-to-noise of the stacked spectra lends a more robust characterization of the average LIS absorption properties of the galaxies from each stack. 

We stacked the LzLCS galaxies based on stellar mass, with a low-mass bin ($M_\star\leq10^8\ M_\odot$), an intermediate-mass bin ($10^8-10^9\ M_\odot$), and a high-mass bin ($M_\star \geq10^9\ M_\odot$). 
These mass bins were selected to roughly separate which physical processes dominate due to their changing gravitational potentials. Dwarf galaxies with $M_\star<10^9\ M_\odot$ are particularly susceptible to stellar feedback because of their shallow gravitational potentials \citep[e.g.,][]{collins22}, with the effects of supernova feedback becoming increasingly important with decreasing stellar mass \citep{lazar26}. 
In particular, galaxies with $M_\star\lesssim10^8\ M_\odot$ can experience strongly time-variable star formation and efficient feedback-driven outflows \citep[e.g.,][]{sparre17,muratov15}, which can preferentially remove metals from their shallow potential wells \citep[e.g.,][]{christensen18}.
At higher stellar masses, deeper gravitational potentials reduce the ability of stellar feedback to globally perturb or expel the ISM, while additional internal processes become increasingly important. A stellar mass of $M_\star \sim10^9\ M_\odot$ marks an approximate transition above which internal processes can begin to dominate \citep[e.g.,][]{lazar26}. We include an intermediate-mass bin to isolate galaxies spanning this transition rather than imposing a simple low- versus high-mass dichotomy. In particular, galaxies with $10^8\ M_\odot<M_\star<10^9\ M_\odot$ occupy a regime in which their gravitational potentials are sufficiently deep to reduce the effectiveness of stellar feedback relative to the lowest-mass dwarfs, while remaining below the mass scale where additional internal regulatory processes become increasingly important \citep[e.g.,][]{lazar26,collins22}. These mass intervals therefore approximately span a progression from shallow-potential systems with highly stochastic star formation and efficient feedback-driven metal loss, through intermediate-mass dwarfs, to more deeply bound systems in which stellar feedback is less effective at globally restructuring the ISM \citep[e.g.,][]{muratov15,christensen18}.

Using physically motivated mass boundaries also facilitates comparison with previous LIS absorption studies and theoretical predictions that often report results in similar mass ranges. It allows us to examine how absorption-line properties change across distinct physical regimes rather than purely statistical divisions of the sample. After restricting our sample to only contain those with uninterrupted spectral coverage of \ion{Si}{2} \W1260 and \ion{C}{2} \W1334, both uncontaminated by Milky Way lines, our final sample consists of 58 LzLCS+ galaxies, for which 9 are in the low-mass bin, 29 in the intermediate, and 20 are high-mass. 

Before constructing these stacks, each spectral region was continuum-normalized using median continuum measurements on both sides of the absorption feature. For the \ion{Si}{2} \W1260 and \ion{C}{2} \W1334 lines, continuum windows (\ion{Si}{2}: 1250 $-$ 1270 \AA, \ion{C}{2}: 1325 $-$ 1340 \AA) were selected sufficiently far from the line center to avoid contamination from the absorption wings while remaining close enough to represent the local continuum level. Each spectrum was divided by the median flux in these continuum windows prior to stacking.

The choice of continuum normalization method can influence the measured absorption depth, particularly in spectral regions with lower signal-to-noise. This is especially relevant for the C II \W1334 region, where the COS sensitivity decreases and continuum noise increases. As discussed by \citet{flury25}, local normalization can introduce additional uncertainty when the continuum level is poorly constrained, potentially affecting derived quantities such as EW and residual flux ($R_f$). However, adopting local continuum windows remains necessary to mitigate large-scale spectral shape variations across the sample.

To maximize the S/N and recover representative low-ionization absorption features, we produced composite spectra using three independent stacking approaches: mean, median, and weighted average stacking. 
In the mean stacks, each normalized spectrum contributes equally to the average, which enhances strong features but can be biased by a small number of galaxies with deep absorption or emission features. 
The median stack, in contrast, is less sensitive to such outliers, yielding a more robust representation of the typical galaxy in each stellar-mass bin. 
Finally, the weighted average stack combines spectra according to their individual continuum S/N, emphasizing higher-quality data while potentially down-weighting faint sources that may still carry physically relevant information. The uncertainties of each stack is derived by propagating the individual uncertainty arrays based on the stack method chosen.

Table~\ref{table:stacking numbers} lists the median properties of the LzLCS galaxies in each bin ($M_\star$, SFR, and $f_{esc}$) and the S/N of the stacked continuum for each stacking method. The final Si~II~$\lambda1260$ and C~II~$\lambda1334$ stacks used in our analysis are shown in Figures~\ref{fig:SiII Stack} and~\ref{fig:CII Stack}, respectively. These figures illustrate the stacked absorption for each stellar-mass bin (corresponding to the columns) and for the three different stacking methods (mean: pink, median: green, weighted average: cyan). The top three rows of these figures correspond to these stacking methods, including their uncertainties (shading of the same color), while the bottom row shows all three on the same figure. The systemic wavelength for each line is denoted by a dashed red line.

Overall, we find that the stacking method chosen has little impact on the final Si~II~$\lambda1260$ profiles. For the C~II~$\lambda1334$, we observe differences in the weighted average stack which shows deeper and slightly broader absorption profile, likely due to the presence of single galaxy with a higher S/N spectra with more pronounced profiles. This is a reminder that stacking spectra from multiple galaxies is a non-trivial process that can be biased from multiple effects. First, when galaxies with different absorption and emission characteristics are averaged together, the resulting stacked spectrum can produce intermediate line strengths that do not necessarily correspond to any individual system. For example, \citet{jaskot19} showed that stacking galaxies with strong Si II absorption but weak Si II* emission together with systems exhibiting the opposite behavior can yield a composite spectrum with moderate absorption and emission in both lines. Further, small velocity offsets can arise from redshift uncertainties or intrinsic galaxy-to-galaxy velocity differences, which can then lead to line broadening and slight misalignment of absorption features. These kinematic effects cause the resulting composite profiles to appear smoother and shallower than any individual spectrum, as narrow components are blended into a single, averaged line shape. 
Emission filling and asymmetric wings may also emerge in the stacks due to overlapping outflow geometries or residual redshift scatter, and the moderate spectral resolution of the LzLCS$+$ data limits our ability to resolve the impact of these narrow kinematic components. Finally, the non-Gaussian COS line spread function further broadens spectral features, redistributing flux from the line core into the wings. As a result, a significant fraction of the intrinsic velocity structure may be smoothed out prior to stacking, such that the additional broadening introduced by stacking represents a secondary effect.  We discuss the impact of these caveats further when discussing the results in Section~\ref{sec4}

Figures~\ref{fig:SiII Stack} and~\ref{fig:CII Stack} show clear trends across the stellar mass sequence for both the median and weighted-average stacks. The intermediate- and high-mass bins exhibit well-defined absorption troughs that deepen and extend modestly in velocity with increasing stellar mass. In contrast, the low-mass bin shows no clearly defined absorption trough in either \ion{Si}{2} or \ion{C}{2}. Instead, the stacked profiles remain close to the continuum level with significant scatter, indicating either intrinsically weak absorption, low covering fractions, or that any absorption present is washed out by noise and stacking effects.
This qualitative difference suggests a transition from poorly constrained or intrinsically weak absorption in low-mass systems to increasingly coherent and structured absorption in more massive galaxies.
In comparison, the intermediate- and high-mass bin stacks show a slight increase in velocity extent of the profiles with mass. This follows the expected scaling relations of outflows with galaxy mass. For example, \cite{xu22} found that the more massive CLASSY galaxies generally host stronger and faster warm outflows. This diversity of profiles makes it interesting in the context of this work to explore whether simulated spectra from a single galaxy can reproduce this line profile diversity. 

In this work, we do not favor a single stacking approach and consider all three throughout our comparison with the RHD simulation. This comparison is discussed more in-depth in Section~\ref{sec4}. In the next section, we present the RHD simulation.

\subsection{RHD Simulation}\label{sec3}

To investigate the physics behind the ultraviolet absorption and emission features of \ion{Si}{2} and \ion{C}{2} observed in the LzLCS+ sample, we adopt the methodology of \citet{gazagnes23, gazagnes24} and compare to synthetic spectra using outputs from the high-resolution RHD simulation presented in \citet{mauerhofer21}.
We briefly summarize here. 
The simulation tracks the formation and evolution of a $\sim10^9\ M_\odot$ galaxy within a cosmological zoom-in framework followed over the redshift range $z = 3.00 - 4.19$, capturing multi-phase ISM physics, star formation, and stellar feedback using the physics from \texttt{SPHINX} \citep{rosdahl18}.
The simulation reaches a maximum cell resolution of approximately 14 pc near z=3, enabling it to resolve dense star-forming structures, diffuse gas, and ionized regions within the ISM. 
Star particles in the simulation emit radiation based on age and metallicity, with spectra derived from the \texttt{BPASS} v2.0 stellar population synthesis models \citep{eldridge08, stanway16}. 
The radiation is propagated using the \texttt{RAMSES-RT} adaptive mesh refinement code \citep{teyssier02, rosdahl13, rosdahl15}, which self-consistently couples radiation and hydrodynamics.

The simulation was post-processed using 75 time step snapshots from z = 3.00–4.19
($\approx 9.2$ Myr time steps) of the simulation using the Monte Carlo radiative transfer code \texttt{RASCAS}, which follows the resonant scattering of UV photons \citep[see][for more details]{mauerhofer21}. 
The radiative-transfer calculation includes dust absorption following the prescription adopted by \citet{gazagnes23}. The dust optical depth in each simulation cell is calculated using the implementation of \citet{katz22}, motivated by the metallicity-dependent dust-to-gas relation of \citet{remyruyer14}. This prescription adopts a broken power-law dependence on metallicity and sets the dust opacity to zero in gas with temperatures above $10^5$K. We also account for the depletion of gas-phase carbon and silicon onto dust grains using metallicity-dependent depletion fractions based on \citet{decia16} and \citet{konstantopoulou23}.
The \ion{Si}{2}~$\lambda1260$ and \ion{C}{2}~$\lambda1334$ LIS absorption lines and their fluorescent emission counterparts, \ion{Si}{2}*~$\lambda1265$ and \ion{C}{2}*~$\lambda1335$, are modeled by simulating the absorption and re-emission of photons through the clumpy, turbulent ISM. We note that these fluorescent lines are not detected in all the mass bins for each ion in the LzLCS+ stacks, possibly due to the moderate S/N and low spectral resolution of the spectra. In addition, the observed strength of fine-structure emission can depend on galaxy geometry and aperture effects. Because these lines arise from resonant scattering and re-emission, their observed flux may depend on how much of the scattered emission region falls within the COS aperture. If lower-mass LyC-leaking galaxies are systematically more compact then a larger fraction of the scattered emission could be captured within the aperture, potentially enhancing the observed fine-structure emission relative to more extended systems. In higher-resolution datasets, the strength of these fine-structure lines has been suggested to trace the fraction of resonantly scattered photons that escape through low-optical-depth channels \citep[e.g.,][]{gazagnes23}.  While we do not explicitly test this possibility in this paper, future work combining the LzLCS$+$ spectroscopy with UV imaging and size measurements could help clarify the role of galaxy compactness in shaping the observed \ion{Si}{2}* and \ion{C}{2}* emission.

For each snapshot, 300 unique lines of sight are sampled using the \texttt{HEALPix} \citep{gorski05} tessellation scheme, which ensures an equal-area distribution over the celestial sphere. This yields a total of 22,500 mock spectra, capturing both spatial and temporal variation in the galaxy's ISM.

Each line of sight intersects different structures in the simulated galaxy, including dense star-forming regions, low-density channels cleared by feedback, and cold gas in the halo, allowing us to explore how geometry and local conditions influence the emergent spectral features. 
The synthetic spectra are generated with high spectral resolution ($\approx 10$ km/s), incorporating both thermal and bulk gas motions, including inflows, outflows, and turbulent velocities.
This suite of mock spectra enables a direct, physically grounded comparison with the observed LzLCS+ data. 

The RHD simulation naturally spans a narrower range of galaxy properties than the observed LzLCS+ sample, as shown in Figure~\ref{fig:hist of M, SFR, Z}.
Nevertheless, its stellar mass and gas-phase metallicity overlap with the central portion of the observed distributions, making it a useful physical reference for modeling the \ion{Si}{2} and \ion{C}{2} absorption features considered here. Stellar mass and metallicity are particularly relevant because they influence the available metal column density and dust attenuation that shape these profiles.

The LzLCS$+$ galaxies generally exhibit higher SFRs than the simulation and are elevated relative to the star-forming main sequence. The star formation history of the simulated galaxy is intrinsically bursty rather than smooth, with the SFR varying by a factor of several over timescales of a few hundred Myr (further discussed in Section~\ref{sec3.3}), broadly consistent with the star-formation histories inferred for LzLCS$+$ galaxies themselves.

This burstiness also shapes the simulated galaxy's apparent UV morphology. Although its intrinsic physical extent does not vary strongly with viewing angle, \cite{gazagnes24} found that its UV half-light radius varies from approximately $0.2$ to $2.5$ kpc across different sightlines (the lower bound being limited by the spatial resolution of the synthetic light cube), because along the most compact sightlines the emergent UV luminosity is dominated by a single bright star-forming region rather than by the galaxy as a whole. Since the LzLCS$+$ sample itself is dominated by compact, UV-bright systems, with typical half-light radii of $0.3-0.6$ kpc (Section~\ref{sec2.1}), the compact end of this simulated sightline distribution provides a plausible morphological analog to the observed galaxies, even though the simulated galaxy's global properties do not match the full LzLCS$+$ population.

The simulation also includes a self-consistent propagation of the ionizing photons through radiative transfer. Hence we can derive, for each line of sight, the amount of ionizing photons escaping the galaxy. We use the same approach as \citet{mauerhofer21}: for each of the 300 lines of sight and 75 time steps, we extract mock spectra of the virtual galaxy between 10 \AA\ and 912 \AA, such that each ionizing photon produced encounters an optical depth that is a sum of contributions from H, He, He+, and dust. We determine the total number of ionizing photons going out of the virial radius and divide it by the intrinsic number produced to get $f_{\mathrm{esc}}$. Here we focus on the escape fraction at 900 \AA, obtained by averaging the spectra between 890 \AA\ and 910 \AA, to be consistent with LyC measurements from the literature. In practice, in the virtual galaxy considered here, with the absence of hard ionizing sources, the escape fraction at 900 \AA\ is a reliable proxy of the total escape fraction of ionizing photons \citep{mauerhofer21}.

\section{\texorpdfstring{\ion{C}{2}}{CII} and \texorpdfstring{\ion{Si}{2}}{SiII} Line Profile Comparison}\label{sec4} 

\begin{figure*}[ht]
    \centering
	\includegraphics[width=\linewidth]{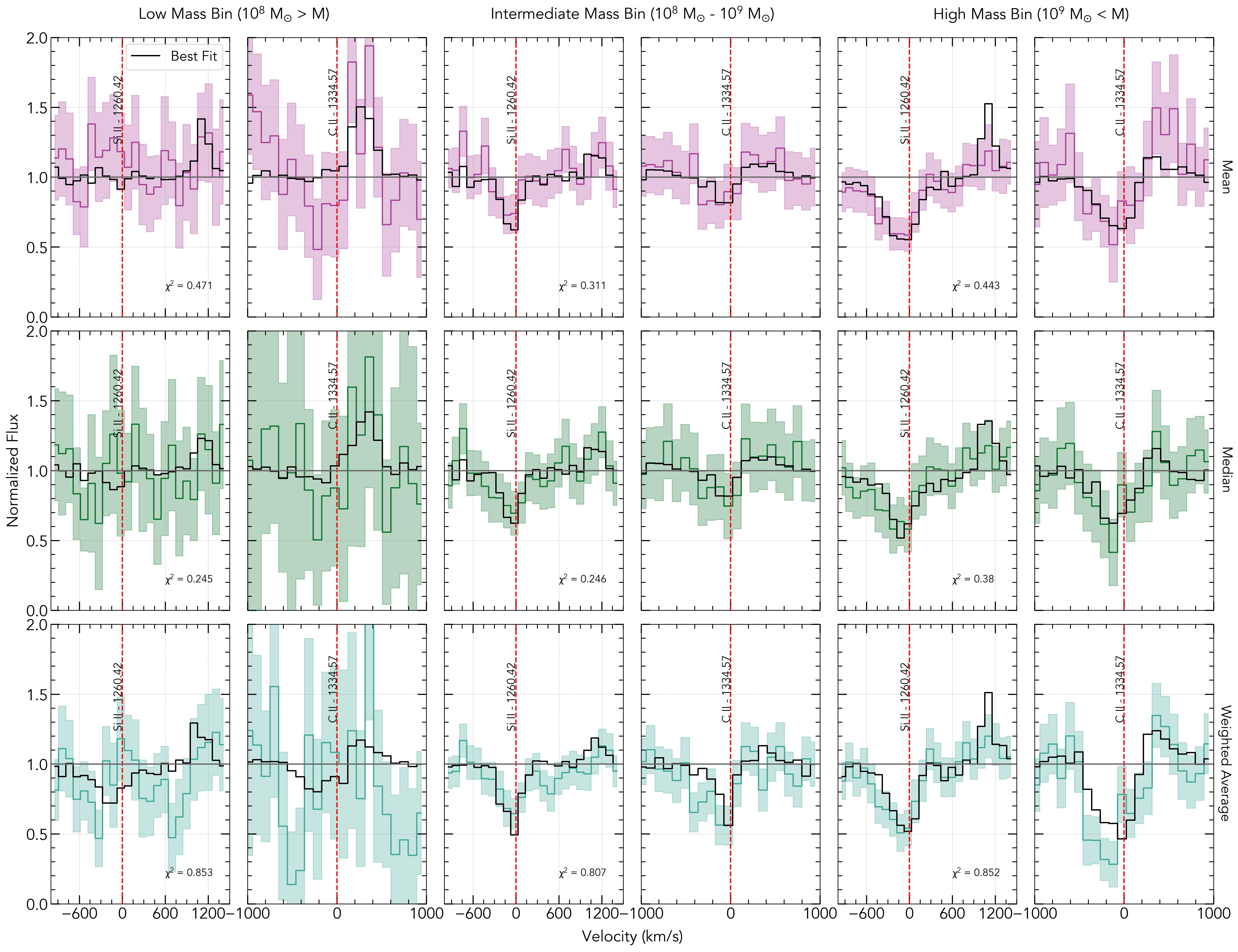}
\caption{Comparison between LzLCS+ stacked spectra and the representative mock spectra drawn from the $\chi^2$-selected subset of 30 best-fitting profiles (black). Rows correspond to stacking method (mean, median, weighted average) and columns to stellar mass bins. Visual inspection combined to the reduced $\chi^2$ values reported in Table~\ref{table:best fit parameters table} indicate reasonable resemblance between simulated profiles and observations. There exist small discrepancies between the observed and simulated spectra in the lowest mass bin, but these differences are difficult to interpret definitively given the limited S/N of these stacks.}
\label{fig:All Stacks w/ Best Fit From Simulation}
\end{figure*}

\begin{deluxetable*}{cCccccc}[ht]
\tablecaption{Best Fit Parameters from RHD Simulation}
\tablehead{
\CH{}     & \CH{$M_\star$ Bin}    & \mcV{Simulation Properties} \\[-1ex] \cline{3-7} 
\CH{Stacking Method} & \CH{($M_\odot$)} & \CH{$\chi^2$} & \CH{Age (Gyr)} & \CH{$Z\ (Z_\odot)$} & \CH{$\log M_\star/M_\odot$} & \CH{$f_{esc}$}}
\startdata
Mean    & < 10^8    & [0.465, 0.503] & 2.00 $\pm$ 0.03 & 0.39 $\pm$ 0.02 & 9.28 $\pm$ 0.06 & 0.19 $\pm$ 0.11 \\
        & 10^8-10^9 & [0.300, 0.452] & 1.96 $\pm$ 0.11 & 0.38 $\pm$ 0.03 & 9.26 $\pm$ 0.07 & 0.22 $\pm$ 0.13 \\
        & > 10^9    & [0.443, 0.563] & 1.87 $\pm$ 0.21 & 0.36 $\pm$ 0.06 & 9.20 $\pm$ 0.14 & 0.09 $\pm$ 0.09 \\
\hline
Median  & < 10^8    & [0.236, 0.254] & 1.98 $\pm$ 0.08 & 0.39 $\pm$ 0.02 & 9.27 $\pm$ 0.05 & 0.19 $\pm$ 0.13 \\
        & 10^8-10^9 & [0.249, 0.342] & 1.97 $\pm$ 0.13 & 0.39 $\pm$ 0.04 & 9.26 $\pm$ 0.09 & 0.14 $\pm$ 0.11 \\
        & > 10^9    & [0.387, 0.494] & 1.92 $\pm$ 0.19 & 0.37 $\pm$ 0.06 & 9.23 $\pm$ 0.12 & 0.09 $\pm$ 0.09 \\
\hline
Weight  & < 10^8    & [0.706, 0.756] & 2.00$\pm$ 0.02 & 0.39 $\pm$ 0.01 & 9.28 $\pm$ 0.01 & 0.19 $\pm$ 0.11 \\
Average & 10^8-10^9 & [0.807, 1.108] & 1.96 $\pm$ 0.13 & 0.38 $\pm$ 0.04 & 9.25 $\pm$ 0.09 & 0.13 $\pm$ 0.10 \\
        & > 10^9    & [0.852, 1.023] & 1.92 $\pm$ 0.18 & 0.37 $\pm$ 0.05 & 9.23 $\pm$ 0.11 & 0.04 $\pm$ 0.05 \\
\enddata
\tablecomments{Best fit parameters from the simulation for the three stacking methods used for LzLCS$+$ in this work. Reported values correspond to the mean properties of the $\sim$30 mock spectra with the lowest $\chi^2$ values for each stack. These values do not represent a unique best-fit solution but rather summarize the range of statistically indistinguishable models consistent with the data.
Column 1 lists the stacking method used for the spectra,
Column 2 lists the three stellar mass bins, and
Column 3 lists the range of $\chi^2$ values for the 30 best matching mock spectra, 
Columns 4-7 lists the mock-derived age, metallicity, stellar mass, and $f_{\rm esc}$ in the simulation. See Section~\ref{sec4} and \ref{sec5} for further details and discussion.
\label{table:best fit parameters table}}
\end{deluxetable*}


The first step this work is to use the mock spectra from the RHD simulation to investigate whether they reproduce of the \ion{Si}{2} \W1260 and \ion{C}{2} \W1334 profiles of the LzLCS+. To compare the stacks to the simulated suite of 22,500 synthetic spectra, we first convolve the simulated spectra to match the resolution of the LzLCS+ observations. We then determine
the best-fit mock spectrum for each stack through a simultaneous $\chi^2$ minimization over the \ion{Si}{2} \W1260 and \ion{C}{2} \W1334 regions, spanning 1256–1265 \AA\ and 1330–1339 \AA, respectively. This joint fitting ensures that both transitions share a consistent best-fit model, constraining the line profiles under the same physical conditions.

We note that, owing to the relatively large uncertainties in the stacked spectra, particularly in the low-mass bin, a wide range of mock spectra produce statistically indistinguishable $\chi^2$ values. Differences such as $\chi^2$ $\approx$ 0.3 versus 0.4 lie well below the 1$\sigma$ confidence threshold and therefore do not represent meaningful differences in fit quality. Consequently, selecting a single best-fit mock spectrum can obscure an existing degeneracy among acceptable simulated sightlines that reproduce the chosen observation. To probe a larger sample of mock spectra that may faithfully reproduce a given observation, we compute the $\chi^2$ values for all 22,500 mock spectra for each stack and examine the full $\chi^2$ distribution. We then follow the methodology of \citet{gazagnes24} whose authors adopt a pragmatic approach by selecting the subset of 30 mock spectra with the lowest $\chi^2$, adopting a maximum threshold of $\chi^2\leq2$. This choice balances the need to sample the degeneracy in acceptable models while avoiding excessive averaging that would mask genuine variation in line profiles. 

We performed a joint fit for all three stacks (median, mean, and weighted average) and extracted the sample of 30 best-fit mock spectra for each. For visualization purposes, Figure~\ref{fig:All Stacks w/ Best Fit From Simulation} compares the observed LzLCS$+$ stacks with the single best-fit mock spectrum (i.e., the one with the lowest $\chi^2$). For each set of 30 mock spectra, we extracted the range of  $\chi^2$ values alongside key physical parameters: the age and metallicity of the virtual galaxy at its original time step, and the escape fraction of ionizing photons along that particular sightline. We report the mean of these parameter distributions as the central value, with uncertainties derived from the standard deviation. These results are summarized in Table \ref{table:best fit parameters table}.

Unless stated otherwise, the simulation-based quantities reported for each observed stack are calculated directly from the $30$ individually selected mock spectra with the lowest $\chi^2$ values. The reported central values and uncertainties therefore summarize the distribution of properties among these acceptable mock sightlines and snapshots; they should not be interpreted as uniquely recovered physical properties of the observed galaxies.

In the following subsections, we visually inspect and discuss the overall line profile agreements (Section~\ref{sec3.1}), and discuss the connection between mock-derived line properties and the galaxy properties  (Section~\ref{sec3.2}), the temporal origin of the best matching profiles (Section~\ref{sec3.3}), and finally compare the $f_{\rm esc}$ escape fractions from the stacks with values estimated using the simulation (Section~\ref{sec3.4}).

\subsection{$\chi^2$, visual inspection and line properties}\label{sec3.1}

Figure~\ref{fig:All Stacks w/ Best Fit From Simulation} shows good overall agreement between the best-matched mock spectra and the empirical stacks. In the lowest-mass bin, the observed \ion{Si}{2} profiles lack clearly defined absorption troughs, nonetheless, the simulation identifies mock spectra that reproduce these features. The observed \ion{C}{2} stacks have larger uncertainties and lower continuum S/N than the corresponding \ion{Si}{2} profiles, making the comparison more difficult to assess. However, these \ion{C}{2} profiles also exhibit more pronounced features, but the high uncertainty in this bin precludes a definitive conclusion regarding the robustness of the fit. In such cases, the profile with significantly higher S/N (here, \ion{Si}{2}) primarily constrains the joint fit.

In the intermediate-mass bin, the simulated lines closely match the observed absorption depth and width. In the high-mass bin, the simulated profiles accurately replicates the absorption troughs in both the mean and median stacks. We note a slight mismatch in the fluorescent emission feature redward of the \ion{Si}{2} absorption in the mean stack, which is notably absent in the median stack. 

In the high-mass weighted-average stack, the observed \ion{C}{2} $\lambda$1334 line is slightly broader and deeper than the best-fit mock spectrum. Interestingly, this discrepancy affects only the \ion{C}{2} profile, as the \ion{Si}{2} line remains well-reproduced. As discussed in Section~\ref{sec2.1}, the weighted-average stack can be biased by a single high-S/N spectrum, which may possess intrinsically deeper absorption at these wavelengths than the average population.

In general, the reduced $\chi^{2} < 1$ values further confirm that the mock spectra provide a close match to the line profiles in the different stacks. While the reduced $\chi^{2}$ is slightly lower for the median stacks and higher for the weighted-average stacks, these offsets likely reflect differences in propagated uncertainties across the stacking methods rather than a statistically superior fit. This highlights how different stacking techniques average down noise in distinct ways, potentially impacting the best model selection, especially in low-S/N regimes where physical inference depends on subtle profile variations.

Previous studies \citep{gazagnes23, gazagnes24} found that while simulated profiles from the same virtual galaxy reproduce LIS lines from a large diversity of galaxy observations with $M_\star < 10^{10}M_\odot$, they tend to fail to represent the deep, broad profiles of the most massive systems. In our analysis, we find that the $\chi^{2}$ distributions for the mean and median stacks extend toward higher values in the high-mass bin, suggesting increased complexity in identifying mock profiles that accurately resemble these massive-system line profiles. However, these differences remain small, and our sample size is insufficient to robustly conclude whether this trend is statistically significant.

\begin{figure*}[ht]
    \centering
	\includegraphics[width=\linewidth]{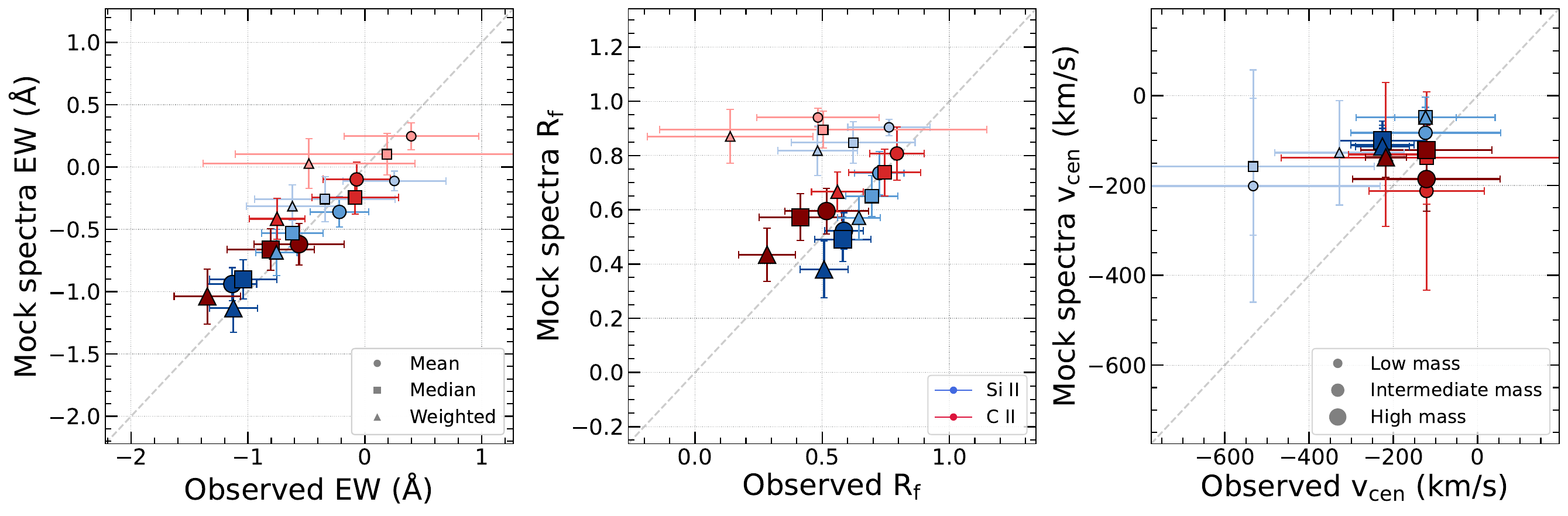}
\caption{Comparisons of simulated and observed absorption line properties for the line EW (left), residual flux (middle), and central outflow velocity (right), across all stacking methods (squares, circles, and triangles), for \ion{Si}{2} (blue) and \ion{C}{2} (red). The points are sized by the mass bin they belong to, with higher-mass stacks shown as larger markers. The simulation-based EW, $R_f$, and $v_{\rm cen}$ are derived from the set of 30 individually selected mock spectra matching a given stacked observation; the reported central values and uncertainties therefore summarize the distribution of properties among these acceptable mock sightlines and snapshots, all of which originate from the same virtual galaxy but at different times and orientations. Note that because mock spectra are selected based on their similarity to the observed line profiles, the agreement between observed and mock EW and $R_f$ is an expected outcome of the profile-matching method rather than an independent validation. The discrepancies seen in the low-mass stack, although accompanied by large uncertainties, indicate that this match is not always one-to-one: $R_f$ and $v_{\rm cen}$ are more sensitive to noise than EW, since they depend on localized flux values near the line center rather than an integral over the full profile, making them more susceptible to being biased by individual noisy pixels in the low-S/N regime.}
\label{fig:comp}
\end{figure*}

To further compare the absorption features in the stacked spectra to the simulation, we measure the EW, the velocity centroid ($v_{\rm cen}$), and the residual flux ($R_f$) for both the \ion{Si}{2} $\lambda1260$ and \ion{C}{2} $\lambda1334$ transitions. All measurements are performed on  the continuum-normalized spectra. The EW is computed by numerically integrating the normalized absorption profile over the wavelength range of the line,
\begin{equation}
{\rm EW} = \int (F_\lambda - 1)\, d\lambda ,
\end{equation}
where $F_\lambda$ is the continuum-normalized flux. In practice, this integral is evaluated using a trapezoidal integration over the pixels that fall within the absorption window.
The velocity centroid is defined as the velocity that bisects the cumulative EW of the absorption profile. Specifically, the cumulative integral of $(F_\lambda-1)$ is computed across the absorption region.

$v_{\rm cen}$ is taken as the equivalent-width-weighted median velocity of the absorption profile. In other words, $v_{\rm cen}$ corresponds to the velocity at which 50\% of the total equivalent width has been accumulated. This definition provides a robust measure of the characteristic velocity of the absorbing gas even when the line profile is asymmetric.

The residual flux is defined as the minimum normalized flux near the line center,
\begin{equation}
R_f = \min(F_\lambda),
\end{equation}
evaluated within a small wavelength window centered on the transition. This quantity traces the depth of the absorption trough and provides an estimate of the covering fraction of the absorbing gas under the assumption of saturated absorption.

Uncertainties on these measurements are estimated using Monte Carlo realizations of the stacked spectra. For each measurement, the flux in every pixel is perturbed according to its associated uncertainty assuming Gaussian noise, and the line properties are recomputed. Repeating this procedure for 1000 realizations yields a distribution of values for EW, $v_{\rm cen}$, and $R_f$, from which the $1\sigma$ uncertainties are taken as the standard deviation of the resulting distributions. The final values and uncertainties are presented in Table \ref{table:properties comparison}.


Figure~\ref{fig:comp} compares the line measurements in the stacks and to those from the best-matching spectra.
We find a strong agreement, but note that discrepancies exist in the low-mass stacks, particularly for the residual flux and central velocities. 
For the former, while the stacks yield low values ($<0.5$), the mock profiles suggest higher values ($>0.8$). 
The observed central outflow velocities are larger by 200 to 400 km s$^{-1}$ compared to the mock spectra. 
This difference is primarily an effect of the lower S/N ratio and absence of clear absorption troughs in the low-mass regime. 
Because $R_f$ and $v_{\rm cen}$ are highly sensitive to localized noise, a single fluctuation can significantly bias the measurement. 
In contrast, fitting simulated profiles is more robust because the model is constrained by flux variations across multiple wavelength bins, effectively averaging out noise features. 
Consequently, extracting line measurements from best-fit mock spectra provides a more reliable method for connecting spectral features to galaxy properties.

Figure~\ref{fig:comp} also highlights that both \ion{Si}{2} and \ion{C}{2} exhibit deeper absorption and larger EWs in more massive stacks. This trend is consistent with an increase in gas covering fraction or column density as stellar mass increases. The scaling relations between line properties and galaxy properties are further discussed in the next section.

\subsection{The link between line profiles and galaxy properties}\label{sec3.2}

\begin{figure*}
    \centering
	\includegraphics[width=\linewidth]{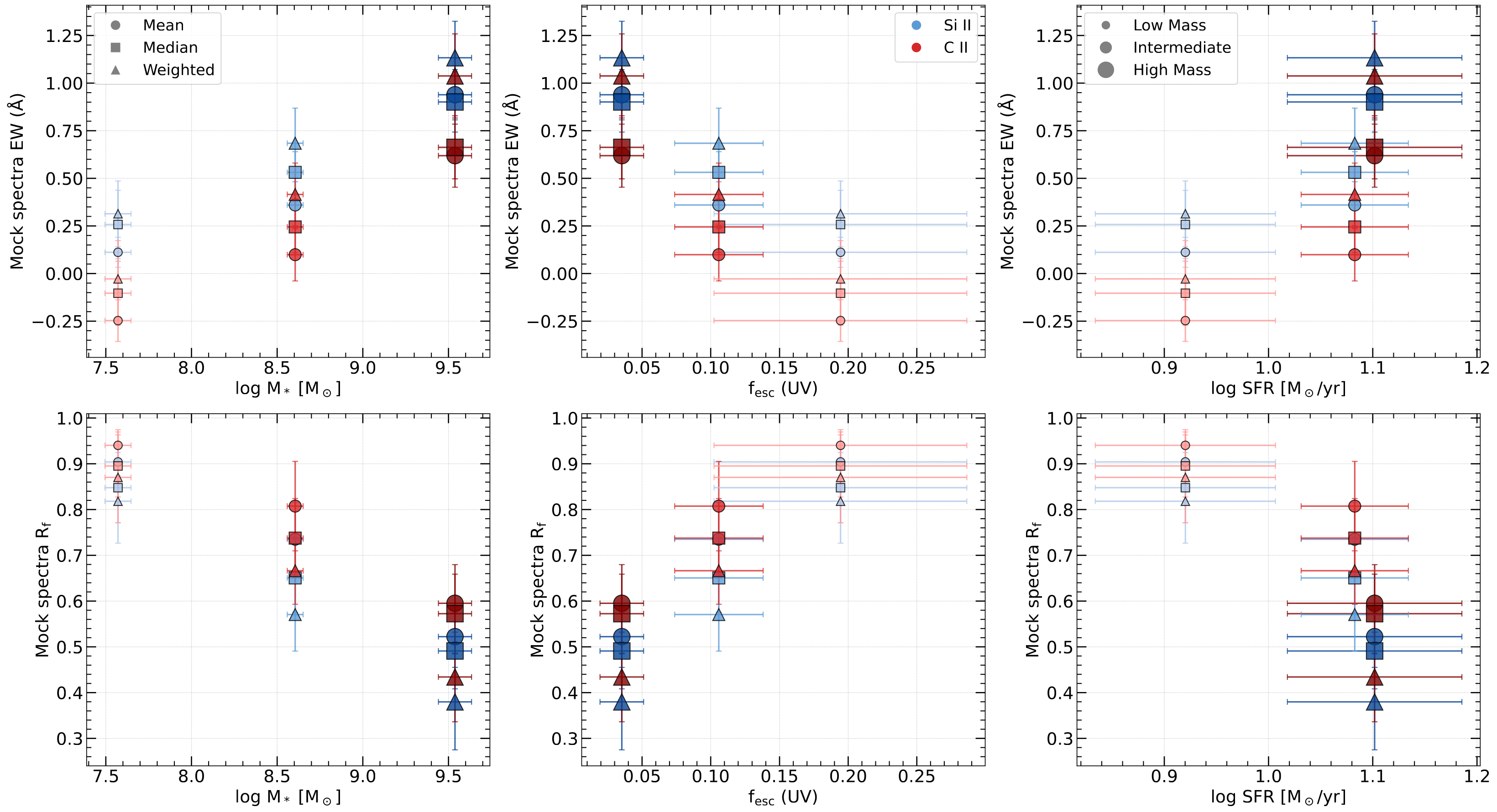}
\caption{Comparison between mock-derived line profile properties (y-axes) and the empirical global parameters of the targeted LzLCS+ stacks (left column: stellar mass; middle column: ionizing escape fraction $f_{\text{esc}}$; right column: star formation rate). Rows display the mock-derived equivalent width (EW, top) and residual flux ($R_f$, bottom). Colors denote different ions (\ion{Si}{2} in blue, \ion{C}{2} in red), shapes represent the stacking method (mean, median, weighted), and marker sizes/opacity correspond to the empirical mass bins. 
The trends between simulated EW and $R_f$ versus observed mass and SFR (left and middle columns) reflect the diversity of gas conditions sampled across sightlines and epochs of the single simulated galaxy. The value of this comparison (see Section~\ref{sec3.2}) lies in showing that empirical scaling relations normally derived from samples of many galaxies are reproduced using sightlines drawn from a single simulated galaxy.}
\label{fig:line_gal}
\end{figure*}

Here we explore how the absorption line properties derived from the mock spectra compare to the average galaxy properties of each stack. Figure~\ref{fig:line_gal} shows that these measurements, specifically the EW and residual flux, exhibit well-defined correlations with stellar mass, SFR, and ionizing escape fraction. These trends are characterized by increasing absorption strength and decreasing residual flux in more massive, higher-SFR systems, leading to smaller measured $f_{\rm esc}$. This directly mirrors the scaling relations and results reported in established LzLCS studies \citep{flury22a, flury22b, flury25, saldana-lopez22}.

We find the trends in  Figure~\ref{fig:line_gal} interesting because they show that these established empirical scaling relations between observed line and galaxy properties are reproduced using line properties drawn entirely from a single $\sim10^9$ galaxy evolution simulation. 
This is a non-trivial result because these scaling relations are normally derived from, and interpreted in the context of, samples of many distinct galaxies spanning a broad range of masses and star-formation histories. Here, the same qualitative behavior emerges purely from the sightline-to-sightline and epoch-to-epoch diversity available within one evolving object, whose own simulated mass barely varies across our comparison, but produces large variations in UV brightness and morphology. 
As shown further in the next section, the sightlines responsible for this diversity are concentrated around specific starburst episodes, suggesting that a single galaxy passing through an active star-forming phase samples a sufficiently broad range of local ISM conditions (i.e. $\sim1.5 - 4.0$ M$_{\odot}$ yr$^{-1}$, $0.20-0.42$ $Z_\odot$) to reproduce trends usually attributed to differences between galaxies.

While the S/N of the individual LzLCS$+$ spectra do not allow for individual galaxy fits, \citet{gazagnes23} performed a similar analysis for the higher S/N spectra of the 45 individual CLASSY galaxies, finding good mock spectra matches for 38 galaxies with the strong trends for the simulated absorption properties and the observed global galaxy properties.
Together, these result suggests that a single, well-resolved RHD simulation can, in principle, be used to explore and help explain the physical origin of population-level scaling relations that are otherwise only accessible empirically through large galaxy samples. Rather than requiring a library of simulated galaxies spanning the full observed mass range to reproduce a given trend, our results indicate that resolving the diversity of local conditions and evolutionary phases within a single object can itself generate much of that trend. This motivates the use of high-resolution, single-galaxy RHD simulations as a complementary tool for isolating which physical mechanisms (e.g., burst-driven changes in covering fraction versus genuine differences in global galaxy mass) underlie observed scaling relations.
This question is difficult to address with observations alone, where mass, SFR, and evolutionary phase all vary simultaneously across a sample.

\subsection{The time origin of the best matching spectra}\label{sec3.3}

\begin{figure}[ht]
    \centering
	\includegraphics[width=\linewidth]{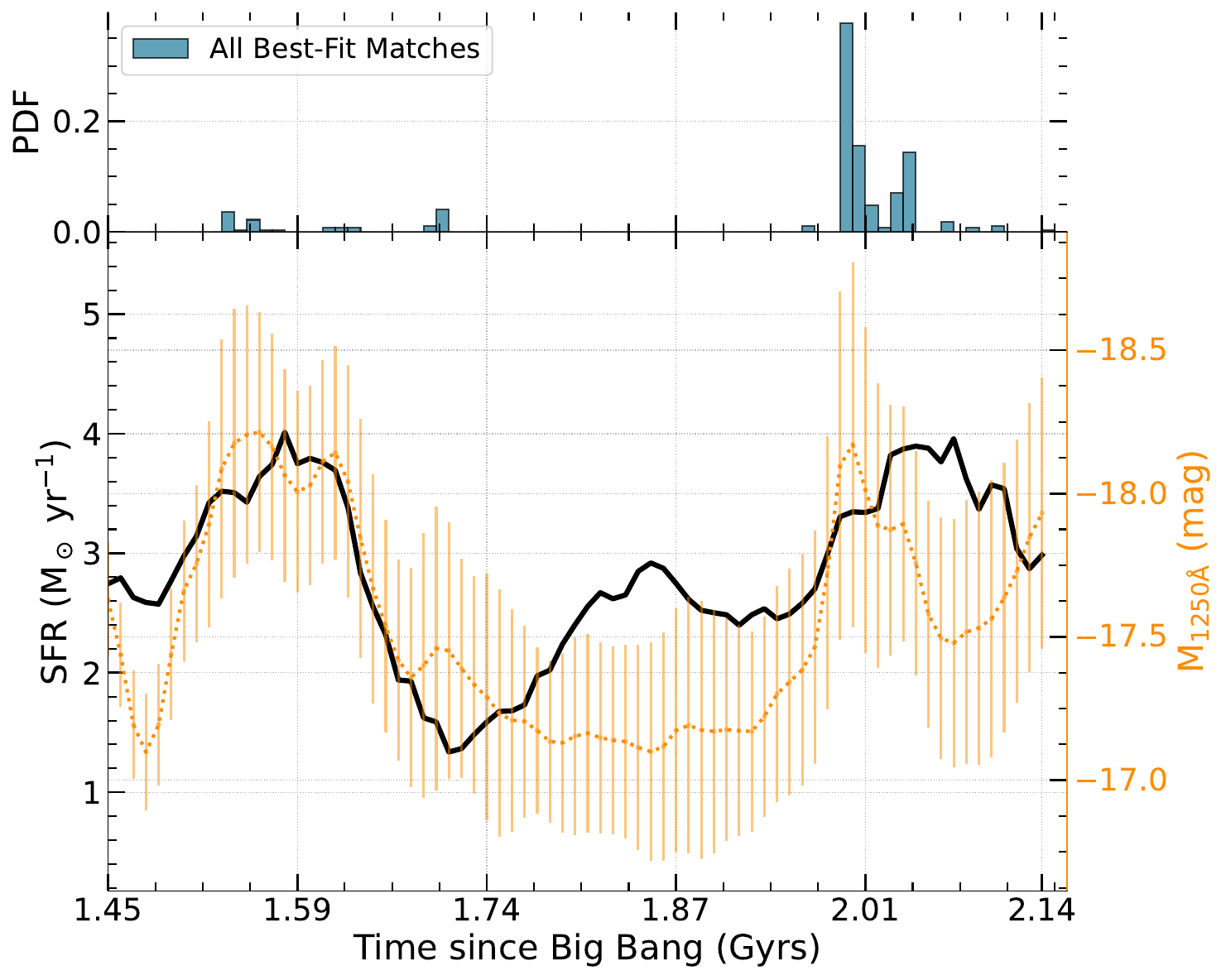}
\caption{The evolution of the SFR (black) and AB magnitude at 1250\AA\ (red) of the virtual galaxy over the 75 time steps used to produce the mock LIS line profiles. The magnitude curve is the median and standard deviation from the 300 line-of-sight measurements at each time step. On the top panel, we present a histogram illustrating the distribution of time steps where the 270 best-matching mock spectra originate (taking the 30 best mocks for each of the 9 stacks).}
\label{fig:sfr}
\end{figure}

The mean properties of the best-matching mock spectra (age, metallicity, stellar mass)  from the best-matching mock spectra (see Table~\ref{table:best fit parameters table}) are tightly constrained over the limited evolutionary track ($\sim$690 Myrs) of the single simulated system we are considering. 
As a result, the dynamic range of these properties is inherently limited, making a direct comparison with the broad property distributions of the LzLCS+ sample primarily illustrative. 
Yet, interestingly, while the simulation covers an evolutionary period from $1.45$ to $2.14$~Gyr, the models that best reproduce the observations for all stacks consistently originate within a narrow temporal window centered around 2~Gyr.

To further investigate, we analyzed the time origin of the 90 best-matching mock spectra (comprising 30 mocks for each of the 3 stacks). In Figure~\ref{fig:sfr}, we present the evolution of the virtual galaxy's star formation rate (SFR) over the timeframe considered in this study. To quantify the UV continuum, we compute the AB magnitude at 1250 \AA\ (M1250) for each simulation output, plotting the median and standard deviation across 300 viewing angles.

Remarkably, most of the best matching mock spectra originate from specific time steps around 2.00 Gyrs, which correspond exactly to the simulation's most extreme SFR peaks. Hence, the temporal origins of the best matching mock spectra are not randomly distributed, but are instead tightly correlated a peak in the virtual galaxy's UV luminosity at 1250 \AA. Accounting for the 100 Myr averaging window used in our star formation rate calculations, this luminosity peak directly trace temporary phases of intense starburst activity.

This aspect was already discussed in \citet{gazagnes24}, when investigating the resemblance of the simulation with the VANDELS sample. This concentration indicates that active starburst phases likely generate a vast diversity of LIS line profiles across different viewing angles. Because star formation governs gas kinematics, and relative changes in the SFR heavily dictate the shape of LIS metal lines, spectra generated during these burst phases cover a sufficient dynamic range to replicate a wide array of varied galaxy observations.

The relatively small number of successful matches near the earlier SFR peak at approximately $1.59$ Gyr suggests that an elevated SFR alone is not sufficient to generate the required diversity of LIS profiles. \citet{gazagnes23} proposed that the first burst may not have been sufficiently energetic to substantially disrupt the surrounding neutral gas, whereas feedback associated with the later burst may have more effectively cleared low-density channels through the ISM. This interpretation was motivated by the lower average metal-line equivalent widths during the second burst. However, equivalent-width evolution alone cannot establish the underlying change in gas structure, and confirming this explanation would require a dedicated analysis of the neutral-gas mass, geometry, and kinematics throughout the simulation. Such an investigation is beyond the scope of the present work.

Ultimately, this result suggests that achieving a robust spectral match does not require the simulated and observed galaxies to share identical masses or large-scale galaxy properties. Rather, it necessitates that the objects occupy a similar evolutionary phase. In this instance, a star-formation burst in the simulation produces a diversity of ISM properties such that the simulated LIS profiles effectively mirrors the range absorption features observed in real star-forming galaxies.

\subsection{Comparing the $f_{\rm esc}$}\label{sec3.4}

\begin{figure}[ht]
    \centering
	\includegraphics[width=\linewidth]{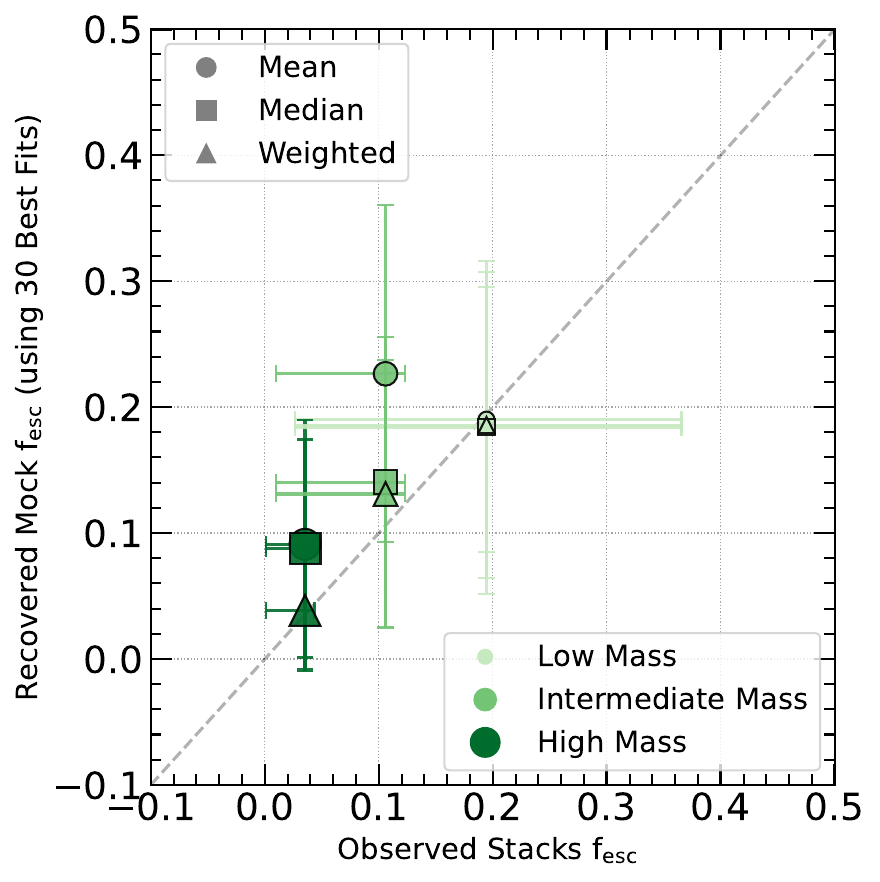}
\caption{Comparison between the observed ionizing escape fraction ($f_{\rm esc}$) from the LzLCS$+$ stacks and the predicted values derived from the 30 best-fit mock spectra. The dashed gray line represents the 1:1 relationship. Symbols denote the stacking method used: circles for mean, squares for median, and triangles for weighted-average stacks. The marker colors and sizes distinguish between the low (light green), intermediate (medium green), and high (dark green) stellar mass bins. Error bars represent the uncertainties in both the empirical measurements and the distribution of the 30 best-fit mock spectra.  There is a  general alignment of the data points with the 1:1 line suggesting that the simulation-based fitting of interstellar absorption lines can provide a robust recovery of the ionizing escape fraction across diverse mass regimes. However, the relatively large uncertainties prevent any strong conclusion.}
\label{fig:fesc}
\end{figure}

Using the strategy detailed in Section~\ref{sec3}, we determine the LyC escape fraction corresponding to each of the 22,500 mock spectra generated. For each stack, we derive a predicted value by calculating the mean and standard deviation of the $f_{esc}$ values from the 30 best-matching mock spectra. To distinguish these simulation-based estimates from direct measurements, we refer to them as $f_{\rm esc}^{\rm virtual}$.

The physical motivation for using the best-matching LIS profiles to estimate $f_{esc}$ is that low-ionization metal absorption and LyC transmission are both regulated by the structure of the intervening neutral gas, including its covering fraction, column density, and velocity distribution. Identifying simulated gas configurations that reproduce the observed \ion{Si}{2} and \ion{C}{2} profiles therefore provides a physically motivated means of examining the range of escape fractions associated with similar line-of-sight gas structures. Nevertheless, the resulting $f_{esc}$ values remain simulation-based predictions and do not constitute direct measurements or unique one-to-one determinations for the observed stacks.

These measurements and their associated uncertainties are reported in Table~\ref{table:best fit parameters table}. Notably, the $f_{\rm esc}^{\rm virtual}$ decreases with increasing stellar mass, mirroring the behavior observed in the stacks. In Figure~\ref{fig:fesc}, we compare the $f_{\rm esc}^{\rm virtual}$ derived from the simulation to the observed stack values reported in Table~\ref{table:stacking numbers}. We find general agreement across stacking methods, although the simulation-derived escape fractions for the mean stack are slightly higher than those for the other methods. In general, the mock spectra matching the lowest stellar mass bins yield the highest $f_{\rm esc}^{\rm virtual}$ values, which are 1-to-1 consistent with the average escape fraction of each stack. These results are consistent with the findings of \citet{gazagnes24}, who showed that escape fractions derived from the same simulation matched direct LyC constraints from the VANDELS survey. This suggests that a simulation-based approach, predicting LyC leakage by matching observed low-ionization state (LIS) profiles to mock spectra, may provide reliable estimates for galaxies lacking direct observations. Furthermore, this method may be more robust than approaches relying on a single line property, as spectral matching remains feasible even at the relatively low S/N and resolution typical of high-redshift observations. The primary role of the present comparison is to establish that the simulated gas configurations and radiative-transfer calculations can generate LIS profiles resembling ensembles of real observations. Future analysis of the resolved gas distribution in the simulation will be required to isolate the causal mechanisms connecting LIS profile morphology to LyC escape.

However, these results should be interpreted with caution, the derived uncertainties are relatively large, and the range of $f_{\rm esc}^{\rm virtual}$ values observed across all simulated sightlines, while spanning the range of the stacks, is ultimately limited, with the maximum value on any single sightline being 47\%.

\begin{deluxetable*}{cccccccccc}[ht]
\tablecaption{Key Parameter Comparison Between Stacks vs. Simulation}
\tablehead{
\CH{} & \CH{Stacking} & \CH{$M_\star$ Bin} & \multicolumn{3}{c}{LzLCS+ Stack Properties} && \multicolumn{3}{c}{Simulation Best Fit Properties} \\[-1ex] \cline{4-6} \cline{8-10}
\CH{Line} & \CH{Method} & \CH{($M_\odot$)} & \CH{$v_{\rm max}$ (km s$^{-1}$)} & \CH{|EW| (\AA)} & \CH{$R_f$} && \CH{$v_{\rm cen}$ (km s$^{-1}$)} & \CH{|EW| (\AA)} & \CH{$R_f$}}
\startdata
\ion{Si}{2} & Mean     & $<10^8    $  & $-530\pm310$ & $-0.25\pm 0.44$  & $0.76\pm0.17$ && $-200\pm150$ & $0.11\pm 0.14$ & $0.90\pm0.04$\\
            &          & $10^8-10^9$ & $-120\pm180$  & $0.22\pm 0.25$ & $0.72\pm0.10$ && $-80\pm60$ & $0.36\pm 0.12$ & $0.74 \pm 0.08$\\
            &          & $>10^9    $  & $-230\pm70$ & $1.13\pm 0.21$  & $0.59\pm0.08$ && $-110\pm40$ & $0.94\pm 0.13$ & $0.52 \pm 0.07$\\
\hline
            & Median   & $<10^8    $ & $-530\pm290$ & $0.34\pm 0.63$ & $0.62\pm0.26$ && $-130\pm180$ & $0.26\pm 0.18$ & $0.85 \pm 0.08$\\
            &          & $10^8-10^9$  & $-120\pm170$ & $0.62\pm 0.26$ & $0.69\pm0.11$ && $-50\pm50$ & $0.53\pm 0.11$ & $0.65 \pm 0.08$\\
            &          & $>10^9    $ & $-230\pm100$ & $1.04\pm 0.28$ & $0.58\pm0.11$ && $-100\pm40$ & $0.90\pm 0.16$ & $0.49 \pm 0.08$\\
\hline 
            & Weighted & $<10^8    $  & $-430\pm130$ & $0.41\pm 0.39$ & $0.63\pm0.14$ && $-130 \pm 180$ & $0.22\pm 0.13$ & $0.88 \pm 0.05$\\
            & Average  & $10^8-10^9$  & $-120\pm80$  & $0.76\pm0.19$ & $0.64\pm0.08$ && $-50\pm 50$ & $0.68\pm 0.18$ & $0.57 \pm 0.08$\\
            &          & $>10^9    $ & $-230\pm60$  & $1.12\pm0.21$ & $0.51\pm0.09$ && $-140\pm40$ & $1.13\pm 0.19$ & $0.38 \pm 0.10$\\
 \hline
 \ion{C}{2} & Mean     & $<10^8    $  & $-220\pm170$& $-0.40\pm 0.61$& $0.48\pm0.34$ && - & $-0.25\pm 0.11$ & $0.94\pm0.04$\\
            &          & $10^8-10^9$ & $-120\pm150$ & $0.07 \pm 0.30$ & $0.79\pm 0.11$ && $-210 \pm 120$ & $0.10 \pm 0.14$ & $0.81\pm0.10$\\
            &          & $>10^9    $  & $-120\pm190$ & $0.56 \pm 0.37$  & $0.52 \pm 0.16$ && $-190\pm70$ & $0.62\pm 0.17$ & $0.60 \pm 0.08$\\
\hline
            & Median   & $<10^8    $  & $-220\pm190$ & $-0.19\pm 1.38$ & $0.50\pm0.67$ && - & -$0.11\pm 0.18$ & $0.90\pm0.07$\\
            &          & $10^8-10^9$  & $-120\pm340$ & $0.08\pm 0.35$ & $0.75\pm0.15$ && $-140\pm100$  & $0.24\pm 0.14$ & $0.74\pm0.09$\\
            &          & $>10^9    $  & $-120\pm180$ & $0.80\pm 0.36$ & $0.41\pm0.17$ && $-120\pm60$ & $0.66\pm 0.18$ & $0.58 \pm 0.09$\\
\hline
            & Weighted & $<10^8    $ & $-510\pm160$ & $0.48\pm0.39$ & $0.14\pm0.61$ && - & $-0.15\pm 0.14$ & $0.91 \pm 0.07$\\
            & Average  & $10^8-10^9$  & $-220\pm90$  & $0.75\pm0.25$ & $0.56\pm0.10$ && $-130\pm180$ & $0.42\pm 0.16$ & $0.67 \pm 0.07$\\
            &          & $>10^9    $  & $-220\pm50$  & $1.35\pm0.30$ & $0.28\pm0.12$ && $-140\pm50$ & $1.04\pm 0.22$ & $0.43 \pm 0.10$\\
\enddata
\tablecomments{Best fit parameters comparison between the simulation and the three stacking methods used for LzLCS$+$.
Column 1 lists the central ion of the stack,
Column 2 lists the stacking method used for the spectra, and
Column 3 lists the three stellar mass bins used.
Columns 4-7 list the LzLCS+ properties that were measured from the stacks: trough velocity ($v_{\rm cen}$), 
EW (positive means absorption dominated), and residual flux ($R_f$). 
In comparison, Columns 7-9 list the trough velocity, EW, and residual flux 
that correspond to the median properties of the $\sim$30 mock spectra fit mock spectra. Dashes for $v_{\rm cen}$ values mean no clear absorption through is present. See Section~\ref{sec3} for further details.
\label{table:properties comparison}}
\end{deluxetable*}


\section{Discussion}\label{sec5}

In this section, we further discuss our results. Section~\ref{sec5.1} discusses the trends between the line profiles and galaxy properties, and Section~\ref{sec5.2} puts this work in the context of two previous efforts comparing this simulation to the RHD Simulation. Section~\ref{sec5.3} connects our findings to high-redshift studies and Section~\ref{sec5.4}  discuss limitations and directions for future work. 

\subsection{Line Profile Trends}\label{sec5.1}

We discuss here briefly how the properties of the LIS lines in observations correlate with physical galaxy parameters. In Figure \ref{fig:line_gal}, we showed a comprehensive comparison of the LzLCS \ion{Si}{2} and \ion{C}{2} absorption line properties derived from the best match mock spectra and how they relate to the stellar mass, SFR, and ionizing photon escape fraction.

Overall, the analysis of the stacked profiles in this work is in line with previous studies. \citet{saldana-lopez22} and \citet{flury25} demonstrate that galaxies with higher ionizing escape fractions ($f_{\rm esc}$) tend to exhibit weaker low-ionization absorption lines, i.e., smaller EWs and higher residual fluxes, consistent with reduced covering fractions or optically thin sightlines through the ISM. Physically, the observed correlation between weaker LIS absorption and higher $f_{\rm esc}$ is often interpreted as arising from reduced neutral gas covering fractions or lower optical depths along LyC escape channels. However, as emphasized by \citet{flury25}, the dominant mechanism depends on the tracer: while \ion{H}{1} absorption primarily reflects variations in covering fraction, metal-line absorption (e.g., \ion{Si}{2}, \ion{C}{2}) can be influenced by both covering fraction and optical depth effects. Consequently, interpreting trends in LIS absorption solely in terms of covering fraction may be oversimplified, and the relationship between absorption strength and $f_{\rm esc}$ likely reflects a combination of geometric and column density variations in the ISM. 

We also observe a clear mass dependence for both the EW and residual flux. The low-mass stack lacks distinct absorption features (exhibiting high residual flux and low EW), suggesting that low-mass galaxies possess highly porous or low-column-density neutral gas distributions whose absorption signatures are too weak to be recovered given the spectral resolution and signal-to-noise ratio of the LzLCS$+$ data. At higher masses, the emergence of well-defined absorption troughs indicates an increasingly structured gas distribution with a higher covering fraction, consistent with more sustained and organized outflows \citep{xu22}. The persistence of these trends across stacking methods supports the conclusion that stellar mass plays a key role in regulating the neutral ISM geometry and kinematics traced by LIS absorption \citep{saldana-lopez22}. Furthermore, as discussed by \citet{flury25}, variations in low-ionization absorption strength can arise from a combination of factors, including neutral gas covering fraction, optical depth, dust attenuation, and feedback-driven ISM structure. Each of these factors is independently known to correlate with stellar mass: dust attenuation increases systematically with $M_\star$ (e.g., \citealt{garn10, whitaker14}), low-ionization/H\,{\sc i} covering fractions and outflow column densities scale with mass and SFR (e.g., \citealt{rubin10,  reddy16, gazagnes18}), and outflow structure becomes more organized in more massive, higher-SFR systems (\citealt{heckman15, xu22}). The mass trend we observe in the LzLCS+ is therefore consistent with these previously established scaling relations, even though disentangling their relative contributions within our own stacks is beyond the scope of this analysis.

Importantly, we must remind that these stacked spectra (and their mock spectra properties) average over significant diversity in galaxy inclination, outflow orientation, and internal ISM kinematics. Some of the velocity extent and residual flux in individual lines could be smeared out by redshift uncertainties or differences in outflow velocities. For example, weighted stacking emphasizes galaxies with high S/N, which can bias the resulting stack toward a subset of galaxies that may not be fully representative of the bin as a whole. On the other hand, median stacking can result in a non-physical spectrum if the individual profiles vary widely within a given bin. Despite these limitations, we find that the relationships between spectral line features and global galaxy properties remain largely independent of the stacking method. The properties derived from the 30 best-matching models reveal consistent physical trends regardless of the stacking technique employed. We note, however, that these findings are specific to the current sample and methodology. Using distinct stacking approaches may introduce different biases in other contexts \citep{flury25}.


\subsection{Comparison with CLASSY and VANDELS}\label{sec5.2}
The comparison between LIS mock spectra from the RHD simulation from \citet{mauerhofer21} and galaxy observations has already been performed in two different context, first with the CLASSY survey \citep{gazagnes23} and then with the VANDELS sample \citep{gazagnes24}. 

While CLASSY and LzLCS exhibit similarly enhanced star formation rates relative to the $z\sim0$ star-forming main sequence, CLASSY galaxies span a significantly broader range of stellar mass ($\log M_\star \sim 6-10$) than the LzLCS$+$ stacks ($\log M_\star \sim 7.5-9.5$). On the other hand, VANDELS is a higher redshift sample ($z>3$) with stronger SFR (10 to 500 $M_\odot$yr$^{-1}$) and higher stellar mass ($\log M_\star \sim 9-10.5$). CLASSY's superior spectral resolution allows for detailed decomposition of kinematic substructures and outflow components that are smoothed out in our lower-resolution, stacked LzLCS$+$ spectra. VANDELS spectra have even lower resolution than the LzLCS (R$\sim600$). Finally, we do not have $f_{\rm esc}$ measurements for CLASSY due to the too low redshift of the sample, but there exists a significant portion of the VANDELS survey with direct $f_{\rm esc}$ constraints.

Our results are in line and consistent with findings from these two previous works. In \cite{gazagnes23} and \cite{gazagnes24}, the authors found surprising high matching accuracy between simulated and observed spectra, with 89\% of the CLASSY galaxies and 83\% of the VANDELS survey being well reproduced by mock spectra from the simulation (using $\chi^2<2$ as threshold). In both studies, the comparison failed for high mass galaxies ($>10^{10}$M$_{\odot}$) that present broad or deep absorption profiles. In this work, although we did observe slightly larger $\chi^2$ values for the high mass stack, these differences are not significant enough to support that the high-mass stack is significantly less well reproduced than the low and intermediate mass counterparts. This may be due to the lower number of very massive galaxies in the LzLCS and to our use of statistical stacking that is at the expense of losing sensitivity to the specific dynamical features of individual massive systems. This tradeoff underscores the complementary nature of detailed, high-resolution studies and population-wide analyses like LzLCS$+$.

As extensively discussed in \citet{gazagnes23, gazagnes24}, the ability of a single simulated galaxy to reproduce line profiles from across distinct mass populations requires careful interpretation. By sampling 300 sightlines over 75 time steps spanning 690~Myr, we capture a diverse range of sightline-dependent ISM conditions in gas density, ionization state, covering fraction, and outflow geometry encountered along different sightlines. Further, the virtual galaxy undergoes two active SFR burst, with the SFR varying between $\sim1.5 - 4.0$ M$_{\odot}$ yr$^{-1}$). The connection between best-matching mock spectra and specific time steps (Section~\ref{sec3.3}) suggests that spectral agreement does not require an exact match in global masses or global galaxy properties. Instead, it may support that the observed and simulated galaxies occupy a similar evolutionary phase, such as an active star-formation burst. This burst generates a diversity of local ISM conditions along different sightlines that effectively replicate the profiles observed in a wide variety of star-forming systems.

\subsection{Implications for High-Redshift Reionization Studies}\label{sec5.3}
At $z > 6$, direct detection of LyC escape is impossible due to the high optical depth of the IGM. As such, UV absorption lines such as \ion{C}{2} and \ion{Si}{2} offer one of the few remaining avenues for indirectly probing the neutral gas content and escape paths of ionizing photons. Our results suggest that even in stacked, low-resolution data, absorption depth and velocity extent retain some sensitivity to the underlying ISM and outflow properties.

Previous studies have established rest-UV absorption lines as diagnostics for Lyman continuum (LyC) escape \citep{heckman11, chisholm2018, steidel2018, saldana-lopez22}, frequently utilizing the gas covering fraction as an empirical proxy for $f_{\rm esc}$ using trends unveiled at low redshift where direct measurements are feasible. This work further supports the utility of these lines for LyC leaking studies at higher redshift. As shown in Section~\ref{sec3.3}, the $f_{\rm esc}$ values derived from the 30  sightlines that best match the \ion{C}{2} and \ion{Si}{2} profiles are consistent with the direct $f_{\rm esc}$ measurements. Notably, our method recovers the empirical trend observed in the LzLCS, where the ionizing escape fraction increases as stellar mass decreases. The mock sightlines identified as best matches for the high-mass stacks consistently yield lower $f_{\rm esc}$ values than those corresponding to the low-mass stacks.

This connection between line strength and escape fraction has a physical origin that can be explored directly within the simulation. Using the same virtual galaxy, \cite{gazagnes24} examined how $f_{\rm esc}^{\rm LyC}$ relates to LIS EW, dust attenuation, and a compactness measure ($r_{75}$, the radius enclosing 75\% of the UV flux) across simulated sightlines, finding that these three properties are strongly correlated: sightlines with low EW(LIS) tend to also have low dust attenuation and small $r_{75}$, and it is this coincidence of properties that produces high $f_{\rm esc}^{\rm LyC}$. Physically, this suggests that LyC photons preferentially escape through spatially compact, dust-poor, and neutral-gas-poor channels, so that weak LIS absorption traces LyC leakage precisely because it is a proxy for the joint absence of neutral gas, dust, and large covering area along a given sightline. We refer the reader to \cite{gazagnes24} for a detailed exploration of this connection.

The use of a simulation based matching approach may be more robust than the use of a single line measurement, and this is because the matching account for the whole line profile, which may extract more information from the combined fit of the absorption and emission features for multiple ions. This matching can be performed and provide relevant results even at low resolution and S/N, assuming the mock spectra are pre-processed to match the instrumental effects affecting the observations.

In practice, however, detecting and characterizing these rest-UV absorption features at $z \geq 6$ requires extremely deep spectroscopy and is often feasible only for the brightest or gravitationally lensed systems. Consequently, applying these diagnostics in a systematic manner across large samples of reionization-era galaxies remains challenging with current facilities. Further, the limitations of stacking methods and the use of single-mass galaxy simulation may also introduce systemic uncertainty. For such diagnostic approach to be applied reliably at high redshift, it is critical to model a broader range of galaxy properties and account for observational and instrumental effects, including resolution and redshift uncertainties. Low-redshift analog samples, such as LzLCS+, remain essential laboratories for calibrating these diagnostics before they can be deployed at the highest redshifts.

\subsection{Limitations of the Current Analysis and Path Forward}\label{sec5.4}

A number of methodological limitations affecting the interpretation of our findings should be reminded. The primary constraint is that stacking procedures can introduce significant biases. Redshift uncertainties, variation in intrinsic profile shapes, and S/N weighting can all wash out or distort absorption features. As noted in \citet{jennings25}, such effects can lead to underestimation of line depths and overestimation of residual flux, particularly in moderate-resolution spectra where saturation is not well-resolved. Overall, that means the properties recovered using a sample of best mock spectra from a RHD simulation may not accurately represent the diversity of properties of the individual objects that make the stacks. 

An additional important caveat is that the RHD simulation is limited by reliance on a single galaxy with $M_\star \sim 10^9 M_\odot$ for all mass bins. 
While we find that this model provides mock spectra that faithfully represent the stacks of low-, intermediate- and high-mass galaxies, the interpretation of such results should be done carefully as one single simulation cannot fully capture the ISM and outflow physics of such a diverse sample of galaxies. It is important to distinguish what local or extreme gas configuration we sample through a extremely fine sampling of time steps and sight lines, and what the global ISM properties of galaxies in diverse mass bin may be. 

Future work should prioritize expanding the library of RHD simulations to cover a range of galaxy masses, metallicities, and star formation histories. This is done for example in the recent work of \citet{mauherhofer26} who produced high-resolution LIS lines for the full 1380 galaxies from the SPHINX20  cosmological radiation-hydrodynamics simulation \citep{rosdahl18, Rosdahl2022}, further demonstrating the potential of LIS lines to predict the LyC escape. Using this latest public release, covering broader mass and galaxy properties would enable more tailored fits to stacked observations and improve interpretation of absorption line trends across bins. Forward modeling approaches that account for instrumental resolution, S/N, and stacking effects are also essential for linking simulations to observations more robustly.

On the observational side, acquiring larger samples of moderate-resolution UV spectra for galaxies at $z \sim 0.5-1$ that are closer analogs to reionization-era systems will help bridge the gap between local high-resolution studies and distant stacked analyses. The continued era of JWST and the upcoming 30-m class telescopes will enable further rest-frame UV spectroscopy of high-redshift galaxies \citep{hammer21}. 
Extending the techniques developed here to such datasets will be vital for constraining the drivers and demographics of LyC escape in the early universe.

\section{Conclusions \& Implications}\label{sec6}

In this work, we compared mock UV \ion{Si}{2} $\lambda$1260 and \ion{C}{2} $\lambda$1334 absorption line profiles, generated from a radiation-hydrodynamics (RHD) simulation of a single $\sim 10^9\ M_\odot$ virtual galaxy, to stacked spectra of 58 galaxies from the LzLCS+ sample. By applying three distinct stacking methodologies (mean, median, and weighted average) across three stellar mass bins  ($M_\star \le 10^8$, $10^8$--$10^9$, and $\ge 10^9\ M_\odot$), we evaluated the ability of synthetic profiles to replicate observed ISM features, investigate possible connections between absorption-line properties and galaxy properties, and provide indirect estimates of ionizing photon escape fractions ($f_{\rm esc}$).

Our primary findings are summarized as follows:

\begin{enumerate}
   \item \textbf{Spectral matching:} For each of the nine stacks (three stacking methods across three mass bins), we identified the 30 mock spectra with the lowest $\chi^{2}$ values. We find overall reasonable agreement between the simulated and observed line profiles, with best-fit mock spectra achieving $\chi^{2} < 1$ across all mass regimes. Notably, we find that extracting line properties such as EW and $R_{f}$ from these best-fit mock profiles provides a relevant alternative to direct empirical measurements. This approach is particularly advantageous in the low-mass, low-S/N regime, where direct measurements are frequently biased by localized noise.
   
    \item \textbf{Scaling relations with galaxy properties:} The LIS mock spectra selected to match the different LzLCS$+$ stellar-mass stacks exhibit systematic trends with the average stellar mass, SFR, and escape fraction of the corresponding observed stacks, broadly reproducing the empirical scaling relations established in previous LzLCS studies \citep{flury22b, saldana-lopez22}. This agreement is notable because all of the mock spectra are drawn from a single virtual galaxy. In particular, the mock spectra show deeper absorption, larger EWs, and lower escape fractions toward higher stellar masses and SFRs, consistent with an increase in the neutral-gas covering fraction and column density.

    \item \textbf{The impact of starburst phases:} We find that the best-matching mock spectra originate almost exclusively from specific time steps corresponding to intense starburst activity and peaks in UV luminosity in the virtual galaxy. This suggests that the diversity of ISM environments required to replicate the line profile observations from a broad range of star-forming galaxies is primarily generated during these active evolutionary phases.
    
    \item \textbf{Recovery of the ionizing escape fraction:} The simulation-based estimates, $f_{\rm esc}^{\rm virtual}$, successfully replicate the observed trend in which more massive galaxies exhibit lower escape fractions. Specifically, the sightlines that provide best-matching mock spectra in the simulation have lower escape fractions for the higher-mass stacks than for the lower-mass bins. Further, the $f_{\rm esc}^{\rm virtual}$ values are 1-to-1 consistent with the average escape fractions of the LzLCS$+$ stacks. However, the relatively large uncertainties associated with these estimates suggest that, while spectral matching is a promising diagnostic for galaxies lacking direct LyC observations, these results must still be interpreted with caution.
\end{enumerate}

In Section~\ref{sec5.3}, we remind that several limitations remain. The current reliance on a single-mass simulation and the inherent biases introduced by stacking highlight the need for a broader range of RHD models and observational studies. Looking ahead, we recommend expanding these efforts to new simulated datasets encompassing a broader range of stellar masses, metallicities, and star formation histories \cite{mauherhofer26}. Doing so would enable more accurate, bin-specific modeling and allow a clearer mapping between physical properties and observable UV line features. Similarly, upcoming observations with JWST and future 30-meter-class ground-based telescopes will provide rest-frame UV spectra of high-redshift galaxies at unprecedented depth. Applying the methodology developed in this study to those datasets will be essential to fully calibrate these simulation-based diagnostics \citep[see also][]{Choustikov2023_, katz2020} and unveil the mechanisms of ionizing photon escape during the Epoch of Reionization.


\begin{acknowledgments}
\nolinenumbers
This research made use of the Low-Redshift Lyman Continuum Survey (\textit{HST}-GO-15626). Additional work was based on observations made with the NASA/ESA Hubble Space Telescope, obtained from the data archive at the Space Telescope Science Institute from HST proposals 13744, 14635, 15341, and 15639. The HST data presented in this article were obtained from the Mikulski Archive for Space Telescopes (MAST) at the Space Telescope Science Institute and can be accessed via \dataset[10.17909/03tp-ak68]{https://doi.org/10.17909/03tp-ak68}.

We acknowledge support from the University of Texas at Austin Department of Astronomy. We also thank the LzLCS and LzLCS$+$ teams for insightful discussions that motivated this analysis.
\end{acknowledgments}

\facilities{HST(COS)}

\software{
Python, 
Astropy \citep{astropy22}, 
NumPy \citep{harris20numpy}, 
Pandas \citep{mckinney10pandas}, 
SciPy \citep{virtanen20scipy}, 
RASCAS \citep{micheldansac20rascas}.}


\typeout{} 
\bibliography{mybib}

@ARTICLE{asplund21,
       author = {{Asplund}, M. and {Amarsi}, A.~M. and {Grevesse}, N.},
        title = "{The chemical make-up of the Sun: A 2020 vision}",
      journal = {\aap},
         year = 2021,
        month = sep,
       volume = {653},
          eid = {A141},
        pages = {A141},
          doi = {10.1051/0004-6361/202140445},
archivePrefix = {arXiv},
       eprint = {2105.01661},
 primaryClass = {astro-ph.SR},
       adsurl = {https://ui.adsabs.harvard.edu/abs/2021A&A...653A.141A}}

@ARTICLE{berg22,
       author = {{Berg}, Danielle A. and {James}, Bethan L. and {King}, Teagan and others},
        title = "{The COS Legacy Archive Spectroscopy Survey (CLASSY) Treasury Atlas}",
      journal = {\apjs},
         year = 2022,
        month = aug,
       volume = {261},
       number = {2},
          eid = {31},
        pages = {31},
          doi = {10.3847/1538-4365/ac6c03},
archivePrefix = {arXiv},
       eprint = {2203.07357},
 primaryClass = {astro-ph.GA},
       adsurl = {https://ui.adsabs.harvard.edu/abs/2022ApJS..261...31B}}

@ARTICLE{james22,
       author = {{James}, Bethan L. and {Berg}, Danielle A. and {King}, Teagan and others},
        title = "{CLASSY. II. A Technical Overview of the COS Legacy Archive Spectroscopic Survey}",
      journal = {\apjs},
         year = 2022,
        month = oct,
       volume = {262},
       number = {2},
          eid = {37},
        pages = {37},
          doi = {10.3847/1538-4365/ac8008},
archivePrefix = {arXiv},
       eprint = {2206.01224},
 primaryClass = {astro-ph.GA},
       adsurl = {https://ui.adsabs.harvard.edu/abs/2022ApJS..262...37J}}

@ARTICLE{flury22a,
       author = {{Flury}, Sophia R. and {Jaskot}, Anne E. and {Ferguson}, Harry C. and others},
        title = "{The Low-redshift Lyman Continuum Survey. I. New, Diverse Local Lyman Continuum Emitters}",
      journal = {\apjs},
         year = 2022,
        month = may,
       volume = {260},
       number = {1},
          eid = {1},
        pages = {1},
          doi = {10.3847/1538-4365/ac5331},
archivePrefix = {arXiv},
       eprint = {2201.11716},
 primaryClass = {astro-ph.GA},
       adsurl = {https://ui.adsabs.harvard.edu/abs/2022ApJS..260....1F}}

@ARTICLE{flury22b,
       author = {{Flury}, Sophia R. and {Jaskot}, Anne E. and {Ferguson}, Harry C. and others},
        title = "{The Low-redshift Lyman Continuum Survey. II. New Insights into LyC Diagnostics}",
      journal = {\apj},
         year = 2022,
        month = may,
       volume = {930},
       number = {2},
          eid = {126},
        pages = {126},
          doi = {10.3847/1538-4357/ac61e4},
archivePrefix = {arXiv},
       eprint = {2203.15649},
 primaryClass = {astro-ph.GA},
       adsurl = {https://ui.adsabs.harvard.edu/abs/2022ApJ...930..126F}}

@ARTICLE{xu22,
       author = {{Xu}, Xinfeng and {Heckman}, Timothy and {Henry}, Alaina and {Berg}, Danielle A. and others},
        title = "{CLASSY III. The Properties of Starburst-driven Warm Ionized Outflows}",
      journal = {\apj},
         year = 2022,
        month = jul,
       volume = {933},
       number = {2},
          eid = {222},
        pages = {222},
          doi = {10.3847/1538-4357/ac6d56},
archivePrefix = {arXiv},
       eprint = {2204.09181},
 primaryClass = {astro-ph.GA},
       adsurl = {https://ui.adsabs.harvard.edu/abs/2022ApJ...933..222X}}

@ARTICLE{jennings25,
       author = {{Jennings}, R. Michael and {Henry}, Alaina and {Mauerhofer}, Valentin and others},
        title = "{A Simulated Galaxy Laboratory: Exploring the Observational Effects on UV Spectral Absorption Line Measurements}",
      journal = {\apj},
         year = 2025,
        month = jan,
       volume = {979},
       number = {1},
          eid = {64},
        pages = {64},
          doi = {10.3847/1538-4357/ad9b13},
archivePrefix = {arXiv},
       eprint = {2412.02794},
 primaryClass = {astro-ph.GA},
       adsurl = {https://ui.adsabs.harvard.edu/abs/2025ApJ...979...64J}}

@ARTICLE{garn10,
       author = {{Garn}, Timothy and {Best}, Philip N.},
        title = "{Predicting dust extinction from the stellar mass of a galaxy}",
      journal = {\mnras},
         year = 2010,
        month = nov,
       volume = {409},
       number = {1},
        pages = {421-432},
          doi = {10.1111/j.1365-2966.2010.17321.x},
archivePrefix = {arXiv},
       eprint = {1007.1145},
 primaryClass = {astro-ph.GA},
       adsurl = {https://ui.adsabs.harvard.edu/abs/2010MNRAS.409..421G}
}

@ARTICLE{whitaker14,
       author = {{Whitaker}, Katherine E. and {Franx}, Marijn and {Leja}, Joel and {van Dokkum}, Pieter G. and {Henry}, Alaina and {Skelton}, Rosalind E. and {Fumagalli}, Mattia and {Momcheva}, Ivelina G. and {Brammer}, Gabriel B. and {Labb{\'e}}, Ivo and {Nelson}, Erica J. and {Rigby}, Jane R.},
        title = "{Constraining the Low-mass Slope of the Star Formation Sequence at 0.5 < z < 2.5}",
      journal = {\apj},
         year = 2014,
        month = nov,
       volume = {795},
       number = {2},
          eid = {104},
        pages = {104},
          doi = {10.1088/0004-637X/795/2/104},
archivePrefix = {arXiv},
       eprint = {1407.1843},
 primaryClass = {astro-ph.GA},
       adsurl = {https://ui.adsabs.harvard.edu/abs/2014ApJ...795..104W}
}

@ARTICLE{gazagnes24,
       author = {{Gazagnes}, Simon and {Cullen}, Fergus and {Mauerhofer}, Valentin and {Begley}, Ryan and {Berg}, Danielle and {Blaizot}, Jeremy and {Chisholm}, John and {Garel}, Thibault and {Leclercq}, Floriane and {McLure}, Ross J. and {Verhamme}, Anne},
        title = "{Comparing the VANDELS Sample to a Zoom-in Radiative Hydrodynamical Simulation: Using the Si II and C II Line Spectra as Tracers of Galaxy Evolution and Lyman Continuum Leakage}",
      journal = {\apj},
         year = 2024,
        month = jul,
       volume = {969},
       number = {1},
          eid = {50},
        pages = {50},
          doi = {10.3847/1538-4357/ad47a4},
archivePrefix = {arXiv},
       eprint = {2405.03759},
 primaryClass = {astro-ph.GA},
       adsurl = {https://ui.adsabs.harvard.edu/abs/2024ApJ...969...50G}
}

@ARTICLE{Garilli2021_vandels,
       author = {{Garilli}, B. and {McLure}, R. and {Pentericci}, L. and {Franzetti}, P. and {Gargiulo}, A. and {Carnall}, A. and {Cucciati}, O. and {Iovino}, A. and {Amorin}, R. and {Bolzonella}, M. and {Bongiorno}, A. and {Castellano}, M. and {Cimatti}, A. and {Cirasuolo}, M. and {Cullen}, F. and {Dunlop}, J. and {Elbaz}, D. and {Finkelstein}, S. and {Fontana}, A. and {Fontanot}, F. and {Fumana}, M. and {Guaita}, L. and {Hartley}, W. and {Jarvis}, M. and {Juneau}, S. and {Maccagni}, D. and {McLeod}, D. and {Nandra}, K. and {Pompei}, E. and {Pozzetti}, L. and {Scodeggio}, M. and {Talia}, M. and {Calabr{\`o}}, A. and {Cresci}, G. and {Fynbo}, J.~P.~U. and {Hathi}, N.~P. and {Hibon}, P. and {Koekemoer}, A.~M. and {Magliocchetti}, M. and {Salvato}, M. and {Vietri}, G. and {Zamorani}, G. and {Almaini}, O. and {Balestra}, I. and {Bardelli}, S. and {Begley}, R. and {Brammer}, G. and {Bell}, E.~F. and {Bowler}, R.~A.~A. and {Brusa}, M. and {Buitrago}, F. and {Caputi}, C. and {Cassata}, P. and {Charlot}, S. and {Citro}, A. and {Cristiani}, S. and {Curtis-Lake}, E. and {Dickinson}, M. and {Fazio}, G. and {Ferguson}, H.~C. and {Fiore}, F. and {Franco}, M. and {Georgakakis}, A. and {Giavalisco}, M. and {Grazian}, A. and {Hamadouche}, M. and {Jung}, I. and {Kim}, S. and {Khusanova}, Y. and {Le F{\`e}vre}, O. and {Longhetti}, M. and {Lotz}, J. and {Mannucci}, F. and {Maltby}, D. and {Matsuoka}, K. and {Mendez-Hernandez}, H. and {Mendez-Abreu}, J. and {Mignoli}, M. and {Moresco}, M. and {Nonino}, M. and {Pannella}, M. and {Papovich}, C. and {Popesso}, P. and {Roberts-Borsani}, G. and {Rosario}, D.~J. and {Saldana-Lopez}, A. and {Santini}, P. and {Saxena}, A. and {Schaerer}, D. and {Schreiber}, C. and {Stark}, D. and {Tasca}, L.~A.~M. and {Thomas}, R. and {Vanzella}, E. and {Wild}, V. and {Williams}, C. and {Zucca}, E.},
        title = "{The VANDELS ESO public spectroscopic survey. Final data release of 2087 spectra and spectroscopic measurements}",
      journal = {\aap},
         year = 2021,
        month = mar,
       volume = {647},
          eid = {A150},
        pages = {A150},
          doi = {10.1051/0004-6361/202040059},
archivePrefix = {arXiv},
       eprint = {2101.07645},
 primaryClass = {astro-ph.GA},
       adsurl = {https://ui.adsabs.harvard.edu/abs/2021A&A...647A.150G}
}

@ARTICLE{McLure2018_vandels,
       author = {{McLure}, R.~J. and {Pentericci}, L. and {Cimatti}, A. and {Dunlop}, J.~S. and {Elbaz}, D. and {Fontana}, A. and {Nandra}, K. and {Amorin}, R. and {Bolzonella}, M. and {Bongiorno}, A. and {Carnall}, A.~C. and {Castellano}, M. and {Cirasuolo}, M. and {Cucciati}, O. and {Cullen}, F. and {De Barros}, S. and {Finkelstein}, S.~L. and {Fontanot}, F. and {Franzetti}, P. and {Fumana}, M. and {Gargiulo}, A. and {Garilli}, B. and {Guaita}, L. and {Hartley}, W.~G. and {Iovino}, A. and {Jarvis}, M.~J. and {Juneau}, S. and {Karman}, W. and {Maccagni}, D. and {Marchi}, F. and {M{\'a}rmol-Queralt{\'o}}, E. and {Pompei}, E. and {Pozzetti}, L. and {Scodeggio}, M. and {Sommariva}, V. and {Talia}, M. and {Almaini}, O. and {Balestra}, I. and {Bardelli}, S. and {Bell}, E.~F. and {Bourne}, N. and {Bowler}, R.~A.~A. and {Brusa}, M. and {Buitrago}, F. and {Caputi}, K.~I. and {Cassata}, P. and {Charlot}, S. and {Citro}, A. and {Cresci}, G. and {Cristiani}, S. and {Curtis-Lake}, E. and {Dickinson}, M. and {Fazio}, G.~G. and {Ferguson}, H.~C. and {Fiore}, F. and {Franco}, M. and {Fynbo}, J.~P.~U. and {Galametz}, A. and {Georgakakis}, A. and {Giavalisco}, M. and {Grazian}, A. and {Hathi}, N.~P. and {Jung}, I. and {Kim}, S. and {Koekemoer}, A.~M. and {Khusanova}, Y. and {Le F{\`e}vre}, O. and {Lotz}, J.~M. and {Mannucci}, F. and {Maltby}, D.~T. and {Matsuoka}, K. and {McLeod}, D.~J. and {Mendez-Hernandez}, H. and {Mendez-Abreu}, J. and {Mignoli}, M. and {Moresco}, M. and {Mortlock}, A. and {Nonino}, M. and {Pannella}, M. and {Papovich}, C. and {Popesso}, P. and {Rosario}, D.~P. and {Salvato}, M. and {Santini}, P. and {Schaerer}, D. and {Schreiber}, C. and {Stark}, D.~P. and {Tasca}, L.~A.~M. and {Thomas}, R. and {Treu}, T. and {Vanzella}, E. and {Wild}, V. and {Williams}, C.~C. and {Zamorani}, G. and {Zucca}, E.},
        title = "{The VANDELS ESO public spectroscopic survey}",
      journal = {\mnras},
         year = 2018,
        month = sep,
       volume = {479},
       number = {1},
        pages = {25-42},
          doi = {10.1093/mnras/sty1213},
archivePrefix = {arXiv},
       eprint = {1803.07414},
 primaryClass = {astro-ph.GA},
       adsurl = {https://ui.adsabs.harvard.edu/abs/2018MNRAS.479...25M}
}

@ARTICLE{Pentericci2018_vandels,
       author = {{Pentericci}, L. and {McLure}, R.~J. and {Garilli}, B. and {Cucciati}, O. and {Franzetti}, P. and {Iovino}, A. and {Amorin}, R. and {Bolzonella}, M. and {Bongiorno}, A. and {Carnall}, A.~C. and {Castellano}, M. and {Cimatti}, A. and {Cirasuolo}, M. and {Cullen}, F. and {De Barros}, S. and {Dunlop}, J.~S. and {Elbaz}, D. and {Finkelstein}, S.~L. and {Fontana}, A. and {Fontanot}, F. and {Fumana}, M. and {Gargiulo}, A. and {Guaita}, L. and {Hartley}, W.~G. and {Jarvis}, M.~J. and {Juneau}, S. and {Karman}, W. and {Maccagni}, D. and {Marchi}, F. and {Marmol-Queralto}, E. and {Nandra}, K. and {Pompei}, E. and {Pozzetti}, L. and {Scodeggio}, M. and {Sommariva}, V. and {Talia}, M. and {Almaini}, O. and {Balestra}, I. and {Bardelli}, S. and {Bell}, E.~F. and {Bourne}, N. and {Bowler}, R.~A.~A. and {Brusa}, M. and {Buitrago}, F. and {Caputi}, K.~I. and {Cassata}, P. and {Charlot}, S. and {Citro}, A. and {Cresci}, G. and {Cristiani}, S. and {Curtis-Lake}, E. and {Dickinson}, M. and {Fazio}, G.~G. and {Ferguson}, H.~C. and {Fiore}, F. and {Franco}, M. and {Fynbo}, J.~P.~U. and {Galametz}, A. and {Georgakakis}, A. and {Giavalisco}, M. and {Grazian}, A. and {Hathi}, N.~P. and {Jung}, I. and {Kim}, S. and {Koekemoer}, A.~M. and {Khusanova}, Y. and {Le F{\`e}vre}, O. and {Lotz}, J.~M. and {Mannucci}, F. and {Maltby}, D.~T. and {Matsuoka}, K. and {McLeod}, D.~J. and {Mendez-Hernandez}, H. and {Mendez-Abreu}, J. and {Mignoli}, M. and {Moresco}, M. and {Mortlock}, A. and {Nonino}, M. and {Pannella}, M. and {Papovich}, C. and {Popesso}, P. and {Rosario}, D.~P. and {Salvato}, M. and {Santini}, P. and {Schaerer}, D. and {Schreiber}, C. and {Stark}, D.~P. and {Tasca}, L.~A.~M. and {Thomas}, R. and {Treu}, T. and {Vanzella}, E. and {Wild}, V. and {Williams}, C.~C. and {Zamorani}, G. and {Zucca}, E.},
        title = "{The VANDELS ESO public spectroscopic survey: Observations and first data release}",
      journal = {\aap},
         year = 2018,
        month = sep,
       volume = {616},
          eid = {A174},
        pages = {A174},
          doi = {10.1051/0004-6361/201833047},
archivePrefix = {arXiv},
       eprint = {1803.07373},
 primaryClass = {astro-ph.GA},
       adsurl = {https://ui.adsabs.harvard.edu/abs/2018A&A...616A.174P}
}

@ARTICLE{mauherhofer26,
       author = {{Mauerhofer}, Valentin and {Blaizot}, J{\'e}r{\'e}my and {Garel}, Thibault and {Verhamme}, Anne and {Gazagnes}, Simon and {Kerutt}, Josephine and {Michel-Dansac}, Leo and {Parker}, Kaelee S. and {Rosdahl}, Joakim and {Saldana-Lopez}, Alberto and {Trebitsch}, Maxime and {Kimm}, Taysun and {Ocvirk}, Pierre and {Teyssier}, Romain},
        title = "{The SPHINX public data release. II. Using low-ionisation absorption lines and dust attenuation to predict Lyman continuum escape}",
      journal = {arXiv e-prints},
         year = 2026,
        month = mar,
          eid = {arXiv:2603.17046},
        pages = {arXiv:2603.17046},
          doi = {10.48550/arXiv.2603.17046},
archivePrefix = {arXiv},
       eprint = {2603.17046},
 primaryClass = {astro-ph.GA},
       adsurl = {https://ui.adsabs.harvard.edu/abs/2026arXiv260317046M}
}

@ARTICLE{chisholm2018,
   author = {{Chisholm}, J. and {Gazagnes}, S. and {Schaerer}, D. and {Verhamme}, A. and 
	{Rigby}, J.~R. and {Bayliss}, M. and {Sharon}, K. and {Gladders}, M. and 
	{Dahle}, H.},
    title = "{Accurately predicting the escape fraction of ionizing photons using rest-frame ultraviolet absorption lines}",
  journal = {\aap},
archivePrefix = "arXiv",
   eprint = {1803.03655},
     year = 2018,
    month = aug,
   volume = 616,
      eid = {A30},
    pages = {A30},
      doi = {10.1051/0004-6361/201832758},
   adsurl = {http://adsabs.harvard.edu/abs/2018A%26A...616A..30C}
}

@ARTICLE{Rosdahl2022,
       author = {{Rosdahl}, Joakim and {Blaizot}, J{\'e}r{\'e}my and {Katz}, Harley and {Kimm}, Taysun and {Garel}, Thibault and {Haehnelt}, Martin and {Keating}, Laura C. and {Martin-Alvarez}, Sergio and {Michel-Dansac}, L{\'e}o and {Ocvirk}, Pierre},
        title = "{LyC escape from SPHINX galaxies in the Epoch of Reionization}",
      journal = {\mnras},
         year = 2022,
        month = sep,
       volume = {515},
       number = {2},
        pages = {2386-2414},
          doi = {10.1093/mnras/stac1942},
archivePrefix = {arXiv},
       eprint = {2207.03232},
 primaryClass = {astro-ph.GA},
       adsurl = {https://ui.adsabs.harvard.edu/abs/2022MNRAS.515.2386R}
}

@ARTICLE{katz2020,
       author = {{Katz}, Harley and {{\v{D}}urov{\v{c}}{\'\i}kov{\'a}}, Dominika and
         {Kimm}, Taysun and {Rosdahl}, Joki and {Blaizot}, Jeremy and
         {Haehnelt}, Martin G. and {Devriendt}, Julien and {Slyz}, Adrianne and
         {Ellis}, Richard and {Laporte}, Nicolas},
        title = "{New methods for identifying Lyman continuum leakers and reionization-epoch analogues}",
      journal = {\mnras},
         year = 2020,
        month = aug,
       volume = {498},
       number = {1},
        pages = {164-180},
          doi = {10.1093/mnras/staa2355},
archivePrefix = {arXiv},
       eprint = {2005.01734},
 primaryClass = {astro-ph.GA},
       adsurl = {https://ui.adsabs.harvard.edu/abs/2020MNRAS.498..164K}
}

@ARTICLE{Choustikov2023_,
       author = {{Choustikov}, Nicholas and {Katz}, Harley and {Saxena}, Aayush and {Cameron}, Alex J. and {Devriendt}, Julien and {Slyz}, Adrianne and {Rosdahl}, Joki and {Blaizot}, Jeremy and {Michel-Dansac}, Leo},
        title = "{The Physics of Indirect Estimators of Lyman Continuum Escape and their Application to High-Redshift JWST Galaxies}",
      journal = {arXiv e-prints},
         year = 2023,
        month = apr,
          eid = {arXiv:2304.08526},
        pages = {arXiv:2304.08526},
          doi = {10.48550/arXiv.2304.08526},
archivePrefix = {arXiv},
       eprint = {2304.08526},
 primaryClass = {astro-ph.GA},
       adsurl = {https://ui.adsabs.harvard.edu/abs/2023arXiv230408526C}
}

@ARTICLE{steidel2018,
       author = {{Steidel}, Charles C. and {Bogosavljevi{\'c}}, Milan and {Shapley},
        Alice E. and {Reddy}, Naveen A. and {Rudie}, Gwen C. and
        {Pettini}, Max and {Trainor}, Ryan F. and {Strom}, Allison L.},
        title = "{The Keck Lyman Continuum Spectroscopic Survey (KLCS): The Emergent Ionizing Spectrum of Galaxies at z~{\ensuremath{\sim}}~3}",
      journal = {\apj},
         year = 2018,
        month = Dec,
       volume = {869},
          eid = {123},
        pages = {123},
          doi = {10.3847/1538-4357/aaed28},
archivePrefix = {arXiv},
       eprint = {1805.06071},
 primaryClass = {astro-ph.GA},
       adsurl = {https://ui.adsabs.harvard.edu/\#abs/2018ApJ...869..123S}
}

@ARTICLE{parker24,
       author = {{Parker}, Kaelee S. and {Berg}, Danielle A. and {Gazagnes}, Simon and others},
        title = "{CLASSY. XI. Tracing Neutral Gas Properties Using UV Absorption Lines and 21 cm Observations}",
      journal = {\apj},
         year = 2024,
        month = dec,
       volume = {977},
       number = {1},
          eid = {104},
        pages = {104},
          doi = {10.3847/1538-4357/ad87cd},
archivePrefix = {arXiv},
       eprint = {2410.00236},
 primaryClass = {astro-ph.GA},
       adsurl = {https://ui.adsabs.harvard.edu/abs/2024ApJ...977..104P}}

@article{xu23,
  author       = {Xu, Xinfeng and Heckman, Timothy and Henry, Alaina and Berg, Danielle A. and others},
  title        = {CLASSY. VI. The Density, Structure, and Size of Absorption-line Outflows in Starburst Galaxies},
  journal      = {The Astrophysical Journal},
  year         = {2023},
  volume       = {948},
  number       = {1},
  pages        = {28},
  doi          = {10.3847/1538-4357/acbf46},
  note         = {Bibcode: 2023ApJ...948...28X}}

@ARTICLE{flury25,
       author = {{Flury}, Sophia R. and {Jaskot}, Anne E. and {Saldana-Lopez}, Alberto and others},
        title = "{The Low-redshift Lyman Continuum Survey: The Roles of Stellar Feedback and Interstellar Medium Geometry in LyC Escape}",
      journal = {\apj},
         year = 2025,
        month = may,
       volume = {985},
       number = {1},
          eid = {128},
        pages = {128},
          doi = {10.3847/1538-4357/adc305},
archivePrefix = {arXiv},
       eprint = {2409.12118},
 primaryClass = {astro-ph.GA},
       adsurl = {https://ui.adsabs.harvard.edu/abs/2025ApJ...985..128F}
}

@ARTICLE{carr23,
       author = {{Carr}, Christopher and {Bryan}, Greg L. and {Fielding}, Drummond B. and {Pandya}, Viraj and {Somerville}, Rachel S.},
        title = "{Regulation of Star Formation by a Hot Circumgalactic Medium}",
      journal = {\apj},
         year = {2023},
        month = may,
       volume = {949},
        pages = {21},
          doi = {10.3847/1538-4357/acc4c7},
archivePrefix = {arXiv},
       eprint = {2211.05115},
 primaryClass = {astro-ph.GA},
       adsurl = {https://ui.adsabs.harvard.edu/abs/2023ApJ...949...21C}}

@ARTICLE{inoue14,
   author = {{Inoue}, A.~K. and {Shimizu}, I. and {Iwata}, I. and {Tanaka}, M.},
    title = "{An updated analytic model for the attenuation by the intergalactic medium}",
  journal = {\mnras},
     year = 2014,
   volume = 442,
    pages = {1805-1820},
      doi = {10.1093/mnras/stu936}}

@ARTICLE{robertson15,
   author = {{Robertson}, B.~E. and {Ellis}, R.~S. and {Furlanetto}, S.~R. and {Dunlop}, J.~S.},
    title = "{Cosmic Reionization and Early Star-forming Galaxies}",
  journal = {\apjl},
     year = 2015,
   volume = 802,
      eid = {L19},
    pages = {L19},
      doi = {10.1088/2041-8205/802/2/L19}
}

@ARTICLE{finkelstein19,
   author = {{Finkelstein}, S.~L. and {D’Aloisio}, A. and {Paardekooper}, J.-P. and others},
    title = "{Conditions for Reionizing the Universe with a Low Galaxy Ionizing Photon Escape Fraction}",
  journal = {\apj},
     year = 2019,
   volume = 879,
      eid = {36},
    pages = {36},
      doi = {10.3847/1538-4357/ab1ea8}}

@ARTICLE{heckman11,
   author = {{Heckman}, T.~M. and {Borthakur}, S. and {Overzier}, R. and others},
    title = "{Extreme Feedback and the Epoch of Reionization: Clues in the Local Universe}",
  journal = {\apj},
     year = 2011,
   volume = 730,
      eid = {5},
    pages = {5},
      doi = {10.1088/0004-637X/730/1/5}
}

@ARTICLE{mauerhofer21,
  author       = {Mauerhofer, V.},
  title        = {UV absorption lines and their potential for tracing neutral hydrogen in distant galaxies},
  journal      = {\textit{Astronomy \& Astrophysics}},
  year         = {2021},
  volume       = {646},
  pages        = {A80},
  doi          = {10.1051/0004-6361/202141452},
  note         = {Bibcode: 2021A\&A...646A..80M}
}

@ARTICLE{reddy16,
       author = {{Reddy}, Naveen A. and {Steidel}, Charles C. and {Pettini}, Max and {Bogosavljevi{\'c}}, Milan and {Shapley}, Alice E.},
        title = "{The Connection Between Reddening, Gas Covering Fraction, and the Escape of Ionizing Radiation at High Redshift}",
      journal = {\apj},
         year = 2016,
        month = sep,
       volume = {828},
       number = {2},
          eid = {108},
        pages = {108},
          doi = {10.3847/0004-637X/828/2/108},
archivePrefix = {arXiv},
       eprint = {1606.03452},
 primaryClass = {astro-ph.GA},
       adsurl = {https://ui.adsabs.harvard.edu/abs/2016ApJ...828..108R}
}

@ARTICLE{gazagnes18,
       author = {{Gazagnes}, S. and {Chisholm}, J. and {Schaerer}, D. and {Verhamme}, A. and {Rigby}, J.~R. and {Bayliss}, M.},
        title = "{Neutral gas properties of Lyman continuum emitting galaxies: Column densities and covering fractions from UV absorption lines}",
      journal = {\aap},
         year = 2018,
        month = aug,
       volume = {616},
          eid = {A29},
        pages = {A29},
          doi = {10.1051/0004-6361/201832759},
archivePrefix = {arXiv},
       eprint = {1802.06378},
 primaryClass = {astro-ph.GA},
       adsurl = {https://ui.adsabs.harvard.edu/abs/2018A&A...616A..29G}
}

@ARTICLE{rubin10,
       author = {{Rubin}, Kate H.~R. and {Weiner}, Benjamin J. and {Koo}, David C. and {Martin}, Crystal L. and {Prochaska}, J. Xavier and {Coil}, Alison L. and {Newman}, Jeffrey A.},
        title = "{The Persistence of Cool Galactic Winds in High Stellar Mass Galaxies between z \raisebox{-0.5ex}\textasciitilde 1.4 and \raisebox{-0.5ex}\textasciitilde1}",
      journal = {\apj},
         year = 2010,
        month = aug,
       volume = {719},
       number = {2},
        pages = {1503-1525},
          doi = {10.1088/0004-637X/719/2/1503},
archivePrefix = {arXiv},
       eprint = {0912.2343},
 primaryClass = {astro-ph.CO},
       adsurl = {https://ui.adsabs.harvard.edu/abs/2010ApJ...719.1503R}
}

@ARTICLE{heckman15,
       author = {{Heckman}, Timothy M. and {Alexandroff}, Rachel M. and {Borthakur}, Sanchayeeta and {Overzier}, Roderik and {Leitherer}, Claus},
        title = "{The Systematic Properties of the Warm Phase of Starburst-Driven Galactic Winds}",
      journal = {\apj},
         year = 2015,
        month = aug,
       volume = {809},
       number = {2},
          eid = {147},
        pages = {147},
          doi = {10.1088/0004-637X/809/2/147},
archivePrefix = {arXiv},
       eprint = {1507.05622},
 primaryClass = {astro-ph.GA},
       adsurl = {https://ui.adsabs.harvard.edu/abs/2015ApJ...809..147H}
}

@ARTICLE{rosdahl18,
  author       = {Rosdahl, Joakim and Katz, Harley and Blaizot, J\'er\'emy and Kimm, Taysun and Michel-Dansac, L\'eo and Garel, Thibault and Haehnelt, Martin and Ocvirk, Pierre and Teyssier, Romain},
  title        = {The SPHINX cosmological simulations of the first billion years: the impact of binary stars on reionization},
  journal      = {Monthly Notices of the Royal Astronomical Society},
  year         = {2018},
  volume       = {479},
  number       = {1},
  pages        = {994--1016},
  doi          = {10.1093/mnras/sty1655},
  note         = {Bibcode: 2018MNRAS.479..994R; also arXiv:1801.07259}
}

@ARTICLE{eldridge08,
  author       = {Eldridge, J. J. and Izzard, R. G. and Tout, C. A.},
  title        = {The effect of massive binaries on stellar populations and supernova rates},
  journal      = {Monthly Notices of the Royal Astronomical Society},
  year         = {2008},
  volume       = {384},
  number       = {4},
  pages        = {1109--1118},
  doi          = {10.1111/j.1365-2966.2007.12796.x},
  note         = {Bibcode: 2008MNRAS.384.1109E}
}

@ARTICLE{stanway16,
  author       = {Stanway, E. R. and Eldridge, J. J. and Becker, G. D.},
  title        = {Stellar population effects on the inferred photon density at high redshift},
  journal      = {Monthly Notices of the Royal Astronomical Society},
  year         = {2016},
  volume       = {456},
  number       = {1},
  pages        = {485--499},
  doi          = {10.1093/mnras/stv2661},
  note         = {Bibcode: 2016MNRAS.456..485S}
}

@ARTICLE{teyssier02,
  author       = {Teyssier, R.},
  title        = {Cosmological hydrodynamics with adaptive mesh refinement},
  journal      = {Astronomy \& Astrophysics},
  year         = {2002},
  volume       = {385},
  pages        = {337--364},
  doi          = {10.1051/0004-6361:20015993},
  note         = {Bibcode: 2002A\&A...385..337T}
}

@ARTICLE{rosdahl13,
  author       = {Rosdahl, J. and Teyssier, R. and Agertz, O.},
  title        = {RAMSES-RT: radiation hydrodynamics in the cosmological context},
  journal      = {Monthly Notices of the Royal Astronomical Society},
  year         = {2013},
  volume       = {436},
  number       = {3},
  pages        = {2188--2231},
  doi          = {10.1093/mnras/stt1722},
  note         = {Bibcode: 2013MNRAS.436.2188R}
}

@ARTICLE{rosdahl15,
  author       = {Rosdahl, Joakim and Schaye, Joop and Teyssier, Romain and Agertz, Oscar},
  title        = {Galaxies that shine: radiation-hydrodynamical simulations of disc galaxies},
  journal      = {Monthly Notices of the Royal Astronomical Society},
  year         = {2015},
  volume       = {451},
  pages        = {34--58},
  doi          = {10.1093/mnras/stv937},
  note         = {Bibcode: 2015MNRAS.451...34R}
}

@article{gorski05,
  author       = {G{\'o}rski, K. M. and Hivon, E. and Banday, A. J. and Wandelt, B. D. and Hansen, F. K. and Reinecke, M. and Bartelmann, M.},
  title        = {HEALPix: A Framework for High‐Resolution Discretization and Fast Analysis of Data Distributed on the Sphere},
  journal      = {The Astrophysical Journal},
  year         = {2005},
  volume       = {622},
  pages        = {759--771},
  doi          = {10.1086/427976},
  note         = {Bibcode: 2005ApJ...622..759G; arXiv:astro-ph/0409513}}

@ARTICLE{gazagnes23,
       author = {{Gazagnes}, Simon and {Mauerhofer}, Valentin and {Berg}, Danielle A. and others},
        title = "{Interpreting the Si II and C II Line Spectra from the COS Legacy Archive Spectroscopic SurveY Using a Virtual Galaxy from a High-resolution Radiation-hydrodynamic Simulation}",
      journal = {\apj},
         year = 2023,
        month = aug,
       volume = {952},
       number = {2},
          eid = {164},
        pages = {164},
          doi = {10.3847/1538-4357/acda2c},
archivePrefix = {arXiv},
       eprint = {2305.19177},
 primaryClass = {astro-ph.GA},
       adsurl = {https://ui.adsabs.harvard.edu/abs/2023ApJ...952..164G}}

@ARTICLE{kimm14,
   author = {{Kimm}, T. and {Cen}, R.},
    title = "{ESCAPE FRACTION OF IONIZING PHOTONS DURING REIONIZATION: EFFECTS DUE TO SUPERNOVA FEEDBACK AND RUNAWAY OB STARS}",
  journal = {\apj},
     year = 2014,
   volume = 788,
      eid = {121},
    pages = {121},
      doi = {10.1088/0004-637X/788/2/121}
}

@ARTICLE{izotov16b,
   author = {{Izotov}, Y.~I. and {Orlitová}, I. and {Schaerer}, D. and others},
    title = "{Detection of high Lyman continuum leakage from four low-redshift compact star-forming galaxies}",
  journal = {\nat},
     year = 2016,
   volume = 529,
    pages = {178-180},
      doi = {10.1093/mnras/stw1205}
}

@ARTICLE{izotov18b,
   author = {{Izotov}, Y.~I. and {Worseck}, G. and {Schaerer}, D. and others},
    title = "{Low-redshift Lyman continuum leaking galaxies with high [O III]/[O II] ratios}",
  journal = {\mnras},
     year = 2018,
   volume = 478,
    pages = {4851-4875},
      doi = {10.1093/mnras/sty1378}
}

@ARTICLE{izotov21,
   author = {{Izotov}, Y.~I. and {Guseva}, N.~G. and {Fricke}, K.~J. and {Henkel}, C.},
    title = "{Lyman continuum leakage from low-mass galaxies with M⋆ < 10^8 M⊙}",
  journal = {\aap},
     year = 2021,
   volume = 646,
      eid = {A138},
    pages = {A138},
      doi = {10.1093/mnras/stab612}
}

@ARTICLE{astropy22,
  author        = {{Astropy Collaboration} and Price-Whelan, A. M. and Lim, P. L. and Earl, N. and Stansby, D. and Bradley, L. and others},
  title         = "{The Astropy Project: Sustaining and Growing a Community-oriented Open-source Project and the Latest Major Release (v5.0) of the Core Package}",
  journal       = {The Astrophysical Journal},
  volume        = {935},
  number        = {2},
  pages         = {167},
  year          = {2022},
  doi           = {10.3847/1538-4357/ac7c74},
  adsurl        = {https://ui.adsabs.harvard.edu/abs/2022ApJ...935..167A},
  archivePrefix = {arXiv},
  eprint        = {2206.14220}
}

@ARTICLE{harris20numpy,
  author  = {Harris, Charles R. and Millman, K. Jarrod and van der Walt, Stéfan J. and Gommers, Ralf and Virtanen, Pauli and Cournapeau, David and Wieser, Eric and Taylor, Julian and Berg, Sebastian and Smith, Nathaniel J. and Kern, Robert and Picus, Matti and Hoyer, Stephan and van Kerkwijk, Marten H. and Brett, Matthew and Haldane, Allan and del Río, Jaime Fernández and Wiebe, Mark and Peterson, Pearu and Gérard-Marchant, Pierre and Sheppard, Kevin and Reddy, Tyler and Weckesser, Warren and Abbasi, Hameer and Gohlke, Christoph and Oliphant, Travis E.},
  title   = {Array programming with {NumPy}},
  journal = {Nature},
  volume  = {585},
  number  = {7825},
  pages   = {357--362},
  year    = {2020},
  doi     = {10.1038/s41586-020-2649-2}
}

@INPROCEEDINGS{mckinney10pandas,
  author    = {McKinney, Wes},
  title     = {Data Structures for Statistical Computing in Python},
  booktitle = {Proceedings of the 9th Python in Science Conference},
  editor    = {van der Walt, Stéfan and Millman, Jarrod},
  pages     = {56--61},
  year      = {2010},
  address   = {Austin, TX},
  url       = {https://pandas.pydata.org/}
}

@ARTICLE{virtanen20scipy,
  author  = {Virtanen, Pauli and Gommers, Ralf and Oliphant, Travis E. and Haberland, Matt and Reddy, Tyler and Cournapeau, David and Burovski, Evgeni and Peterson, Pearu and Weckesser, Warren and Bright, Jonathan and {van der Walt}, Stéfan J. and Brett, Matthew and Wilson, Joshua and Millman, K. Jarrod and Mayorov, Nikolay and Nelson, Andrew R. J. and Jones, Eric and Kern, Robert and Larson, Eric and Carey, CJ and Polat, İbrahim and Feng, Yu and Moore, Eric W. and VanderPlas, Jake and Laxalde, Denis and Perktold, Josef and Cimrman, Robert and Henriksen, Ian and Quintero, E. A. and Harris, Charles R. and Archibald, Anne M. and Ribeiro, Antonio H. and Pedregosa, Fabian and {van Mulbregt}, Paul and SciPy 1.0 Contributors},
  title   = {SciPy 1.0: Fundamental Algorithms for Scientific Computing in Python},
  journal = {Nature Methods},
  volume  = {17},
  number  = {3},
  pages   = {261--272},
  year    = {2020},
  doi     = {10.1038/s41592-019-0686-2}
}

@ARTICLE{micheldansac20rascas,
  author  = {Michel-Dansac, L. and Blaizot, J. and Garel, T. and Verhamme, A. and Kimm, T. and Trebitsch, M.},
  title   = {RASCAS: RAdiation SCattering in Astrophysical Simulations},
  journal = {Astronomy \& Astrophysics},
  volume  = {635},
  pages   = {A154},
  year    = {2020},
  doi     = {10.1051/0004-6361/201834961},
  adsurl  = {https://ui.adsabs.harvard.edu/abs/2020A&A...635A.154M}
}

@ARTICLE{saldana-lopez22,
       author = {{Saldana-Lopez}, Alberto and {Schaerer}, Daniel and {Chisholm}, John and others},
        title = "{The Low-Redshift Lyman Continuum Survey. Unveiling the ISM properties of low-z Lyman-continuum emitters}",
      journal = {\aap},
         year = 2022,
        month = jul,
       volume = {663},
          eid = {A59},
        pages = {A59},
          doi = {10.1051/0004-6361/202141864},
archivePrefix = {arXiv},
       eprint = {2201.11800},
 primaryClass = {astro-ph.GA},
       adsurl = {https://ui.adsabs.harvard.edu/abs/2022A&A...663A..59S}}

@ARTICLE{hammer21,
       author = {{Hammer}, F. and {Morris}, S. and {Cuby}, J.-G. and {Kaper}, L. and {Steinmetz}, M. and {Afonso}, J. and {Barbuy}, B. and {Bergin}, E. and {Finogenov}, A. and {Gallego}, J. and {Kassin}, S. and {Miller}, C. and {{\"O}stlin}, G. and {Pentericci}, L. and {Schaerer}, D. and {Ziegler}, B. and {Chemla}, F. and {Dalton}, G. and {De Frondat}, F. and {Evans}, C. and {Le Mignant}, D. and {Puech}, M. and {Rodrigues}, M. and {Sanchez-Janssen}, R. and {Taburet}, S. and {Tasca}, L. and {Yang}, Y. and {Zanchetta}, S. and {Dohlen}, K. and {Dubbeldam}, M. and {El Hadi}, K. and {Janssen}, A. and {Kelz}, A. and {Larrieu}, M. and {Lewis}, I. and {MacIntosh}, M. and {Morris}, T. and {Navarro}, R. and {Seifert}, W.},
        title = "{MOSAIC on the ELT: High-multiplex Spectroscopy to Unravel the Physics of Stars and Galaxies from the Dark Ages to the Present Day}",
      journal = {The Messenger},
         year = 2021,
        month = mar,
       volume = {182},
        pages = {33-37},
          doi = {10.18727/0722-6691/5220},
archivePrefix = {arXiv},
       eprint = {2011.03549},
 primaryClass = {astro-ph.GA},
       adsurl = {https://ui.adsabs.harvard.edu/abs/2021Msngr.182...33H}}

@ARTICLE{mascia24,
       author = {{Mascia}, S. and {Pentericci}, L. and {Calabr\`o}, A. and {Santini}, P. and {Napolitano}, L. and {Arrabal Haro}, P. and {Castellano}, M. and {Dickinson}, M. and {Ocvirk}, P. and {Lewis}, J. S. W. and {Amor\'in}, R. and {Bagley}, M. and {Cleri}, R. N. J. and {Costantin}, L. and {Dekel}, A. and others},
        title = "{New insight on the nature of cosmic reionizers from the CEERS survey}",
      journal = {\aap},
         year = {2024},
        month = may,
       volume = {685},
        pages = {A3},
          doi = {10.1051/0004-6361/202347884},
archivePrefix = {arXiv},
       eprint = {2309.02219},
 primaryClass = {astro-ph.GA},
       adsurl = {https://ui.adsabs.harvard.edu/abs/2024A&A...685A...3M}}

@ARTICLE{jaskot19,
       author = {{Jaskot}, Anne E. and {Dowd}, Tara and {Oey}, M. S. and {Scarlata}, Claudia and {McKinney}, Jordan},
        title = "{New Insights on Lyα and Lyman Continuum Radiative Transfer in the Greenest Peas}",
      journal = {\apj},
         year = {2019},
        month = nov,
       volume = {885},
        pages = {96},
          doi = {10.3847/1538-4357/ab3d3b},
archivePrefix = {arXiv},
       eprint = {1909.04644},
 primaryClass = {astro-ph.GA},
       adsurl = {https://ui.adsabs.harvard.edu/abs/2019ApJ...885...96J}}

@ARTICLE{dayal24,
       author = {{Dayal}, Pratika and {Volonteri}, Marta and {Greene}, Jenny E. and {Kokorev}, Vasily and {Goulding}, Andy D. and {Williams}, Christina C. and {Furtak}, Lukas J. and {Zitrin}, Adi and {Atek}, Hakim and {Chemerynska}, Iryna and {Feldmann}, Robert and {Glazebrook}, Karl and {Labbe}, Ivo and {Nanayakkara}, Themiya and {Oesch}, Pascal A. and {Weaver}, John R.},
        title = "{UNCOVERing the contribution of black holes to reionization in the JWST era}",
      journal = {arXiv e-prints},
         year = {2024},
        month = jan,
          eid = {arXiv:2401.11242},
        pages = {arXiv:2401.11242},
archivePrefix = {arXiv},
       eprint = {2401.11242},
 primaryClass = {astro-ph.GA},
          doi = {10.48550/arXiv.2401.11242},
       adsurl = {https://ui.adsabs.harvard.edu/abs/2024arXiv240111242D}}

@ARTICLE{grazian24,
       author = {{Grazian}, Andrea and {Giallongo}, Emanuele and {Boutsia}, Kalliopi and {Cristiani}, Stefano and {Vanzella}, Eros and {Guaita}, Lucia and {Nonino}, Mario and {Pentericci}, Laura and {Santini}, Paola and {Castellano}, Marco},
        title = "{What Are the Pillars of Reionization? Revising the AGN Contribution to the Ionizing Background at z > 4}",
      journal = {\apj},
         year = {2024},
        month = nov,
       volume = {974},
        pages = {84},
          doi = {10.3847/1538-4357/ad6980},
archivePrefix = {arXiv},
       eprint = {2407.20861},
 primaryClass = {astro-ph.GA},
       adsurl = {https://ui.adsabs.harvard.edu/abs/2024ApJ...974...84G}}

@ARTICLE{izotov16a,
       author = {{Izotov}, Y. I. and {Orlitov{\'a}}, I. and {Schaerer}, D. and {Thuan}, T. X. and {Verhamme}, A. and {Guseva}, N. G. and {Worseck}, G.},
        title = "{Eight per cent leakage of Lyman continuum photons from a compact, star-forming dwarf galaxy}",
      journal = {\nat},
         year = {2016},
        month = jan,
       volume = {529},
       number = {7585},
        pages = {178-180},
          doi = {10.1038/nature16456},
archivePrefix = {arXiv},
       eprint = {1601.03068},
 primaryClass = {astro-ph.GA},
       adsurl = {https://ui.adsabs.harvard.edu/abs/2016Natur.529..178I}}

@ARTICLE{izotov18a,
       author = {{Izotov}, Y. I. and {Worseck}, G. and {Schaerer}, D. and {Guseva}, N. G. and {Thuan}, T. X. and {Fricke}, K. J. and {Verhamme}, A. and {Orlitov{\'a}}, I.},
        title = "{J1154+2443: a low-redshift compact star-forming galaxy with a 46 per cent leakage of Lyman continuum photons}",
      journal = {\mnras},
         year = {2018},
        month = mar,
       volume = {474},
        pages = {4514-4528},
          doi = {10.1093/mnras/stx3115},
archivePrefix = {arXiv},
       eprint = {1709.05260},
 primaryClass = {astro-ph.GA},
       adsurl = {https://ui.adsabs.harvard.edu/abs/2018MNRAS.474.4514I}}

@MISC{xu24,
       author = {{Xu}, Xinfeng and {Strom}, Allison L. and {Dayal}, Pratika and {Jones}, Tucker and {Stark}, Daniel P. and others},
        title = "{Galactic Winds in the Early Universe: observing outflows in emission and absorption in a typical z~6 galaxy}",
      howpublished = {JWST Cycle 3 General Observer Program 5293},
         year = {2024},
         note = {Space Telescope Science Institute},
          url = {https://www.stsci.edu/jwst/science-execution/program-information?id=5293}}

@ARTICLE{wang19,
       author = {{Wang}, B. and {Heckman}, T. M. and {Amor{\'\i}n}, R. and {Overzier}, R. and {Leitherer}, C. and {Grimes}, J. P. and {Puschnig}, J.},
        title = "{A New Technique for Finding Galaxies Leaking Lyman Continuum Radiation}",
      journal = {\apj},
         year = {2019},
        month = nov,
       volume = {885},
        pages = {57},
          doi = {10.3847/1538-4357/ab4414},
archivePrefix = {arXiv},
       eprint = {1908.06193},
 primaryClass = {astro-ph.GA},
       adsurl = {https://ui.adsabs.harvard.edu/abs/2019ApJ...885...57W}}

@ARTICLE{katz22,
       author = {{Katz}, Harley and {Garel}, Thibault and {Rosdahl}, Joakim and {Mauerhofer}, Valentin and {Kimm}, Taysun and {Blaizot}, J{\'e}r{\'e}my and {Michel-Dansac}, L{\'e}o and {Devriendt}, Julien and {Slyz}, Adrianne and {Haehnelt}, Martin},
        title = "{Mg II in the JWST Era: A Probe of Lyman Continuum Escape?}",
      journal = {\mnras},
         year = 2022,
        month = sep,
       volume = {515},
       number = {3},
        pages = {4265--4286},
          doi = {10.1093/mnras/stac1437},
archivePrefix = {arXiv},
       eprint = {2203.14937},
 primaryClass = {astro-ph.GA},
       adsurl = {https://ui.adsabs.harvard.edu/abs/2022MNRAS.515.4265K}
}

@ARTICLE{remyruyer14,
       author = {{R{\'e}my-Ruyer}, Aur{\'e}lie and {Madden}, Suzanne C. and {Galliano}, Fr{\'e}d{\'e}ric and {Lebouteiller}, Vianney and {Baes}, Maarten and {Bendo}, George J. and {Boquien}, M{\'e}d{\'e}ric and {Boselli}, Alessandro and {Ciesla}, Laure and {Cormier}, D. and {Cortese}, Luca and {De Looze}, Ilse and {Gallagher}, John S. and {Hony}, S. and {Hughes}, Tom M. and {Karczewski}, Olivier L. and {Kirkpatrick}, Allison and {K{\"o}hler}, M. and {O'Halloran}, Barry and {Parkin}, T. J. and {Roussel}, H{\'e}l{\`e}ne and {Wilson}, Christine D.},
        title = "{Gas-to-dust mass ratios in local galaxies over a 2 dex metallicity range}",
      journal = {\aap},
         year = 2014,
        month = mar,
       volume = {563},
          eid = {A31},
        pages = {A31},
          doi = {10.1051/0004-6361/201322803},
archivePrefix = {arXiv},
       eprint = {1312.3442},
 primaryClass = {astro-ph.GA},
       adsurl = {https://ui.adsabs.harvard.edu/abs/2014A%26A...563A..31R}
}

@ARTICLE{decia16,
author = {{De Cia}, Annalisa and {Ledoux}, C{'e}dric and {Mattsson}, Lars and {Petitjean}, Patrick and {Srianand}, Raghunathan and {Gavignaud}, Isabelle and {Jenkins}, Edward B.},
title = "{Dust-depletion sequences in damped Lyman-{\ensuremath{\alpha}} absorbers: A unified picture from low-metallicity systems to the Galaxy}",
journal = {\aap},
year = 2016,
month = dec,
volume = {596},
eid = {A97},
pages = {A97},
doi = {10.1051/0004-6361/201527895},
archivePrefix = {arXiv},
eprint = {1608.08621},
primaryClass = {astro-ph.GA},
adsurl = {https://ui.adsabs.harvard.edu/abs/2016A%26A...596A..97D}
}

@ARTICLE{konstantopoulou23,
       author = {{Konstantopoulou}, Christina and {De Cia}, Annalisa and {Krogager}, Jens-Kristian and {Ledoux}, C{\'e}dric and {Noterdaeme}, Pasquier and {Fynbo}, Johan P. U. and {Heintz}, Kasper E. and {Watson}, Darach and {Andersen}, Anja C. and {Ramburuth-Hurt}, Tanita and {Jermann}, Iris},
        title = "{Dust depletion of metals from local to distant galaxies: I. Peculiar nucleosynthesis effects and grain growth in the ISM (Corrigendum)}",
      journal = {\aap},
         year = 2023,
        month = jun,
       volume = {674},
          eid = {C1},
        pages = {C1},
          doi = {10.1051/0004-6361/202243994e},
archivePrefix = {arXiv},
       eprint = {2208.00878},
 primaryClass = {astro-ph.GA},
       adsurl = {https://ui.adsabs.harvard.edu/abs/2023A%26A...674C...1K}}

@ARTICLE{lazar26,
author = {{Lazar}, Ilin and {Kaviraj}, Sugata and {Martin}, Garreth and {Conselice}, Christopher J. and {Koudmani}, Sophie and {Watkins}, Aaron E. and {Yi}, Sukyoung K. and {Kakkad}, Darshan and {Sedgwick}, Thomas M. and {Dubois}, Yohan and {Devriendt}, Julien E. G. and {Kraljic}, Katarina and {Peirani}, Sebastien},
title = "{Downsizing does not extend to dwarf galaxies: identifying the stellar mass regimes shaped by supernova and AGN feedback}",
journal = {\mnras},
year = {2026},
month = apr,
volume = {547},
number = {2},
pages = {stag207},
doi = {10.1093/mnras/stag207},
archivePrefix = {arXiv},
eprint = {2602.09094},
primaryClass = {astro-ph.GA}}

@ARTICLE{collins22,
author = {{Collins}, Michelle L. M. and {Read}, Justin I.},
title = "{Observational constraints on stellar feedback in dwarf galaxies}",
journal = {Nature Astronomy},
year = {2022},
month = jun,
volume = {6},
pages = {647-658},
doi = {10.1038/s41550-022-01657-4},
archivePrefix = {arXiv},
eprint = {2205.06825},
primaryClass = {astro-ph.GA}}

@ARTICLE{christensen18,
       author = {{Christensen}, Charlotte R. and {Dav{\'e}}, Romeel and {Brooks}, Alyson and {Quinn}, Thomas and {Shen}, Sijing},
        title = "{Tracing Outflowing Metals in Simulations of Dwarf and Spiral Galaxies}",
      journal = {\apj},
         year = 2018,
        month = nov,
       volume = {867},
       number = {2},
          eid = {142},
        pages = {142},
          doi = {10.3847/1538-4357/aae374},
archivePrefix = {arXiv},
       eprint = {1808.07872},
 primaryClass = {astro-ph.GA},
       adsurl = {https://ui.adsabs.harvard.edu/abs/2018ApJ...867..142C}}

@ARTICLE{muratov15,
       author = {{Muratov}, Alexander L. and {Kere{\v{s}}}, Du{\v{s}}an and {Faucher-Gigu{\`e}re}, Claude-Andr{\'e} and {Hopkins}, Philip F. and {Quataert}, Eliot and {Murray}, Norman},
        title = "{Gusty, gaseous flows of FIRE: galactic winds in cosmological simulations with explicit stellar feedback}",
      journal = {\mnras},
         year = 2015,
        month = dec,
       volume = {454},
       number = {3},
        pages = {2691-2713},
          doi = {10.1093/mnras/stv2126},
archivePrefix = {arXiv},
       eprint = {1501.03155},
 primaryClass = {astro-ph.GA},
       adsurl = {https://ui.adsabs.harvard.edu/abs/2015MNRAS.454.2691M}}

@ARTICLE{sparre17,
       author = {{Sparre}, Martin and {Hayward}, Christopher C. and {Feldmann}, Robert and {Faucher-Gigu{\`e}re}, Claude-Andr{\'e} and {Muratov}, Alexander L. and {Kere{\v{s}}}, Du{\v{s}}an and {Hopkins}, Philip F.},
        title = "{(Star)bursts of FIRE: observational signatures of bursty star formation in galaxies}",
      journal = {\mnras},
         year = 2017,
        month = apr,
       volume = {466},
       number = {1},
        pages = {88-104},
          doi = {10.1093/mnras/stw3011},
archivePrefix = {arXiv},
       eprint = {1510.03869},
 primaryClass = {astro-ph.GA},
       adsurl = {https://ui.adsabs.harvard.edu/abs/2017MNRAS.466...88S}}

\end{document}